# Assessing Geographical and Seasonal Influences on Energy Efficiency of Electric Drayage Trucks

Ankur Shiledar[a,b,*], Manfredi Villani[a], Joseph N.E. Lucero[c,d], Ruixiao Sun[e], Vivek A. Sujan[e], Simona Onori[f,d], Giorgio Rizzoni[a,b]

[a] Center for Automotive Research, The Ohio State University, 930 Kinnear Rd, Columbus, OH, 43212

[b] Department of Mechanical and Aerospace Engineering, The Ohio State University, 201 W 19th Ave, Columbus, OH 43210

[c] Department of Chemistry, Stanford University, Stanford, CA 94305

[d] Applied Energy Division, SLAC National Accelerator Laboratory, Menlo Park, CA 94025

[e] Buildings and Transportation Science Division, Oak Ridge National Laboratory, Oak Ridge, TN 37831

[f] Department of Energy Science and Engineering, Stanford, CA 94305

[*] Corresponding author (email: shiledar.1@osu.edu)

## Abstract

The electrification of heavy-duty vehicles is a critical pathway towards improved energy efficiency of the freight sector. The current battery electric truck technology poses several challenges to the operations of commercial vehicles, such as limited driving range, sensitivity to climate conditions, and long recharging times. Estimating the energy consumption of heavy-duty electric trucks is crucial to assess the feasibility of the fleet electrification and its impact on the electric grid. This paper focuses on developing a model-based simulation approach to predict and analyze the energy consumption of drayage trucks used in ports logistic operations, considering seasonal climate variations and geographical characteristics. The paper includes results for three major container ports within the United States, providing region-specific insights into driving range, payload capacity, and charging infrastructure requirements, which will inform decision-makers in integrating electric trucks into the existing drayage operations and plan investments for electric grid development.

This manuscript has been authored by UT-Battelle, LLC under Contract No. DE-AC05-00OR22725 with the U.S. Department of Energy. The United States Government retains and the publisher, by accepting the article for publication, acknowledges that the United States Government retains a non-exclusive, paid-up, irrevocable, world-wide license to publish or reproduce the published form of this manuscript, or allow others to do so, for United States Government purposes. DOE will provide public access to these results of federally sponsored research in accordance with the DOE Public Access Plan (https://www.energy.gov/doe-public-access-plan).

# 1. Introduction

Moving goods is an essential component of the United States economy. Heavy-duty (HD) trucks have the largest share of freight movement, contributing to 45% of the total ton-miles (i.e., weight transported per unit distance), 60% of the total weight of shipments, and 62% of the total value of shipments. In addition, forecasts predict a 50% increase in total freight movement by 2050, with an increasingly larger share of trucks. As such, the trucking industry contributes to 5.6% of the United States (U.S.) gross domestic product (GDP), or $1.3 trillion, with almost 2 million people employed as drivers for heavy-duty and tractor-trailers in 2021 (U.S. Department of Transportation 2022), (U.S. Department of Energy 2023).

The economic predominance of the transportation sector is reflected in its energy demand and related emissions. Globally, transportation is responsible for 23% of emissions from fossil fuel combustion, with the U.S. being one of the largest contributors (U.S. Environmental Protection Agency 2024a). Within the U.S., transportation has, since 2016, become the largest source of greenhouse gas (GHG) emissions, surpassing electricity generation (U.S. Energy Information Administration (EIA) 2016). Therefore, the need to improve the energy efficiency of the freight sector has become apparent. Although light-duty vehicles (LDVs) and passenger cars account for the largest share of GHG emissions in the transportation sector (57% in 2021 (U.S. Environmental Protection Agency 2021)), when examining individual vehicles, HDVs are found to travel six times the annual mileage of LDVs while consuming up to 20 times more fuel (U.S. Environmental Protection Agency 2024b). Therefore, despite HDVs constituting only 5% of the vehicle fleet in the U.S., they are responsible for over a quarter of fuel consumption and transportation emissions.

The electrification of HDVs can have a major role in reducing the energy consumption of transportation and its impact on air quality. The pathway to electrification of the transportation sector will happen in steps and faces many challenges. The transition to 100% sales of zero-emissions vehicles (ZEVs) is expected to take a minimum of two decades for HDV (U.S. Department of Energy 2023). In this process, it is important to identify sub-classes of heavy-duty vehicles or fleets, quantify their energy and operational requirements, and assess whether the current state-of-the-art powertrain technology and infrastructure can support the phased adoption of ZEVs for these sub-classes or fleets, and when. Different fleet types may require additional support to transition to electric trucks (Sugihara, Hardman, and Kurani 2024). Currently, the ZEV technologies include battery electric and hydrogen fuel cell powertrains. The adoption of battery electric trucks (BETs) and fuel cell electric trucks will require significant investments to produce clean electricity and hydrogen, as well as the development of resilient energy distribution networks for recharging and refueling.

In this study, battery electric trucks are considered, and analytical models are developed to accurately quantify their energy requirements. As pointed out in (Hall and Lutsey 2019), since 2019, most of the announced or in production zero emissions heavy-duty trucks are battery electric, with considerably fewer fuel cell options. Furthermore, the study focuses on drayage trucks, a specific sub-class of heavy-duty vehicles. "Drayage" is the transport of freight containers from an ocean port to a destination over short to medium distances (usually below 400 km), commonly referred to as "first mile" in the logistics industry. A pictorial illustration of drayage operations is presented in Figure 1. For a mathematical formulation of the drayage operation problem, see for example (Shiri and Huynh 2018). Drayage loads typically have departure and arrival points in the same region and do not focus on long-haul movements. Nonetheless, drayage is an integral part of the logistics industry and vital to the overall supply chain management process.

The urgency of electrifying drayage trucks stems from their impact on congestion and environmental problems, specifically in metropolitan areas where movement of freight has become a growing concern for the authorities. Environmental impacts of port-related activities and their environmental justice implications have been documented near the Ports of New York (NY) and New Jersey (NJ), with high volumes of truck traffic and higher exposure levels to emissions in low-income neighborhoods. Similarly, the Los Angeles–Long Beach port complex stands as the largest source of diesel pollution, due to drayage truck activity in the surrounding area (Giuliano and O'Brien 2007).

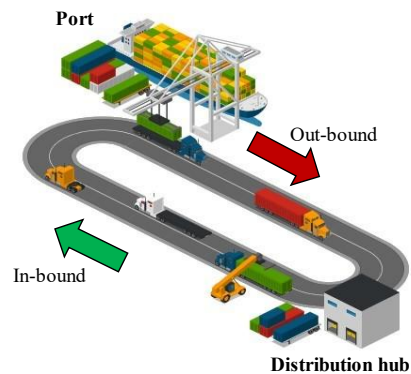

Figure 1: Pictorial representation of drayage operations or "first mile" logistics. Containers are shipped using drayage trucks from an ocean port (origin) to a certain distribution hub (destination) over "out-bound" routes. Trucks then return to the port traveling over "in-bound" routes. Figure created with an open-sourced graph designer (https://icograms.com).

In the U.S., there are several large ports, e.g., Seattle, NY, NJ, Savannah, Houston, Los Angeles, and Long Beach. Each of these ports is in a different region, each with their own local (seasonal) climate variations and geographical characteristics. The energy consumption of ground vehicles strongly depends on the driver's behavior (e.g., acceleration and deceleration aggressiveness or cruising speed) as well as the road and environmental conditions (Demir, Bektaş, and Laporte 2014; Golbasi and Kina 2022). For example, seasonal temperature variations greatly affect the energy requirements for the cabin heating, ventilation, and air conditioning (HVAC) system and other auxiliary systems. This is particularly true for electric vehicles, which cannot rely on the engine's waste heat for cabin heating. Furthermore, very low and very high temperatures can drastically affect the performance of the battery, its instantaneous effective capacity, and its longevity. Exposure to these extreme conditions leads to a reduction in the vehicle's available driving range. In addition, the seasonal temperature variations also affect the resistive drag and rolling resistance forces acting on the vehicle, while the geography of the region, i.e., elevation and road grade, largely affects vehicle energy consumption, with hilly and mountainous regions typically resulting in a higher energy consumption. In sum, the energy requirements for the deployment of electric trucks will vary significantly across regions. This variability will impact not only vehicle design in terms of required on-board energy storage, but also the energy demand to the grid for recharging, and the operations and operating costs of fleets.

This work follows a bottom-up approach to develop comprehensive energy consumption models for electric drayage trucks, with particular emphasis on accounting for driver behaviors, seasonal climate conditions, and geographical factors such as elevation and speed limits, to facilitate region-specific adoption of BETs and inform decision makers on how to plan for their deployment without disruptive effects on the current port operations and logistics. As mentioned earlier, the adoption of BETs in the drayage sector will greatly depend on demonstrating whether electric powertrains and state-of-the-art battery technologies can achieve parity with the current diesel engine technology in terms of fulfilling the freight movement demand from origins to destinations and in terms of cost competitiveness (Giuliano et al. 2021).

Hence, the contribution of this work is to develop simulation tools to analyze the seasonal and regional energy requirements of electric drayage trucks. The development of this simulation-based tools is relevant to many other important research questions pertinent to work being conducted as a part of the undertaken research, such as:

- Understanding the energy consumption variation and resulting impact on driving range and payload capacity of electric drayage trucks in different geographical and weather conditions.
- Understanding the influence of seasonal variation in energy requirements of the truck, and its consideration into recharging/refueling requirements to service the drayage fleet.
- Evaluating the cost-competitiveness of electric trucks in the drayage fleet which will be more rigorous when combined with the detailed vehicle energy consumption models.

This work focuses on developing detailed energy consumption models for drayage trucks, followed by a case study on drayage operations at the Port of Savannah, GA, to investigate the first two points listed above. Similar analysis is conducted for two other major container ports, namely the port of Houston, TX and port of Seattle/Tacoma, WA. These three ports are in distinct regions of the U.S., resulting in stark differences in geography and vastly different weather conditions throughout the year. Considering these characteristically distinct ports allows for a better understanding of the variation in energy consumption experienced by drayage trucks, whether diesel- or battery-powered. Further assessment of the cost-competitiveness of electric drayage trucks at these ports is planned for future studies.

## 2. Related Work

In this section, previous research work is reviewed to show how the proposed models and simulations strengthen the state-of-the-art methods for energy consumption evaluation of electric trucks, and how this work extends the understanding of electrifying the drayage sector. Previous work in this research area focused on the feasibility of drayage electrification, addressing the following research questions:

- Is state-of-the-art battery technology capable of handling the current drayage fleet operations and allow a transition to zero emissions heavy-duty vehicles?
- Can the electric grid support the charging infrastructure?
- What is the overall impact of the deployment of BETs on well-to-wheels emissions?

Tanvir et al., (2020) present a well-structured viability assessment of operating an electric drayage fleet in Southern California. Real-world operational data of the current conventional diesel fleet is collected from 20 trucks. Using a virtual vehicle simulator the electrical energy usage and the battery state of charge for BETs is estimated on the real-world collected data. The analysis explored various scenarios with differing assumptions for battery charging and truck scheduling. The results indicate that 85% of the tours could be completed by electric trucks with a 250 kWh battery capacity if charging opportunities are available at the home base during the intervals between the tours. However, the energy consumption model presented in this study is quite simple and based on several assumptions. Despite the efforts of the authors of considering conservative assumptions, their model mostly uses constant climate conditions, component efficiencies, and auxiliary loads, e.g., HVAC, even though this can greatly affect the energy consumption estimation, reducing (or increasing) the percentage of tours that could be served by the electric trucks, therefore limiting the reliability of the results. Also, the model seems to not account for the geographical characteristics of the region, neglecting the power consumption related to road slope and elevation.

Exhaustive reports on drayage electrification for the Port of New York and New Jersey (Kotz et al. 2022) and on the total cost of ownership for Class 8 tractors (Hunter et al. 2021) have been published by the National Renewable Energy Laboratory (NREL). NREL researchers used the FASTSim model (Brooker et al. 2015) to estimate the energy consumption of BETs and other vehicle technologies. While FASTSim accounts for road grade, auxiliary loads, and powertrain inefficiencies, the models are not explicitly dependent on the seasonal and regional climate conditions, such as ambient temperature, which in the present work are shown to have a significant impact for BETs.

Some studies propose region-specific case studies to analyze BETs energy consumption and feasibility or compare alternative HD powertrain technologies based on well-to-wheels GHG emissions and cost of ownership. For example, Mojtaba Lajevardi et al. (2019) consider British Columbia, Zhang et al. (2022) examine China as a case study, and Middela et al. (2022) focus on India. While these studies factor in the impact of the grid energy mix, battery manufacturing, and recycling, on the final GHG emissions and costs, they also acknowledge the sensitivity of the results to the accuracy of the models used for estimating the energy consumption, which do not include region-specific considerations.

Recent works by Borlaug et al. (2021) and Shoman et al. (2023) have addressed the charging requirements for the electrification of heavy-duty trucks. Both studies rely on assumptions for the average energy consumption of electric trucks and use sensitivity analysis to account for the uncertainty of this parameter. In particular, the former work explores the sensitivity of the results when energy consumption is varied between 1.5 and 2.8 kWh/mile (baseline, 1.8 kWh/mile), which shows that the daily energy requirements vary significantly. The latter work discusses how the average energy consumption for BETs varies significantly among studies with values ranging from 1.2–1.23 kWh/km for estimates for 2030, to real-world driving data showing 0.80–1.20 kWh/km on urban roads and 1.30–1.80 kWh/km on highways.

The need for numerical simulations to evaluate the energy consumption of BETs emerges from the very limited availability of real-world or experimental data. To the best of the Authors' knowledge, one of the few studies reporting BET experimental data is the work by Sato et al. (2022). The purpose of this study is to quantitatively investigate the energy consumption and energy regeneration rates for medium- and heavy-duty battery-powered electric vehicles using different driving cycles on a chassis dynamometer. Therefore, the results from Sato et al. (2022) have been used as a basis for the validation of the models developed in the present work.

Zhang et al. (2021) utilizes a physics-based model for the analysis of the energy consumption in a mid-size passenger EV, with the influence of ambient temperature and average driving speed studied on energy efficiency. However, the equations do not explicitly show the dependance of energy consumption on the ambient temperature. In another study by Bai et al. (2024) to determine the cost-effectiveness of renewable energy pathways to decarbonize heavy-duty trucks in China used a linear relationship between vehicle energy consumption and vehicle weight to estimate the energy requirements of the electrified heavy-duty fleet. Even though the impact of vehicle weight on energy consumption was captured using a simple model, the influence of other geographical factors and weather conditions were excluded.

The presented literature review shows that while addressing the research questions regarding the feasibility of drayage electrification requires accurate energy consumption models, previous studies often overlook seasonal and geographical impacts. Figure 2 presents an overview of the different modeling approaches that can be employed to develop vehicle energy consumption models.

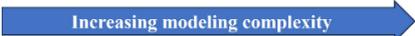

Figure 2: Different approaches available in the literature for determining the vehicle energy consumption in simulations. An increase in modeling level corresponds to greater complexity and longer simulation times. The grey box highlights the most commonly used approaches in feasibility analyses for drayage fleet electrification, while the red box indicates the modeling methods used in this study.

Almost all the works cited above predominantly utilize Level 1 or Level 2 modeling approaches to determine the vehicle energy consumption in their analysis. By relying on conservative assumptions and worst-case scenarios, most researchers can assess the feasibility of drayage trucks electrification, but this can lead to overestimating the actual energy needs of the trucks (i.e., the required capacity of the battery pack) and the fleet's electric energy demand from the grid, potentially hindering the adoption of ZEVs. To address this gap, this study develops model-based vehicle simulators using a digital twin approach to analyze vehicle energy consumption. These simulators incorporate advanced modeling techniques for powertrain and road load components, enabling a detailed evaluation of the significant impact of seasonal and geographical factors on energy consumption. The choice of a Level 2 modeling approach over the detailed methods from Level 3 approaches, was based on the trade-off between model accuracy, complexity and simulation time. The models and simulation study presented here aim to enhance the analysis in existing research and provide valuable insights to stakeholders for planning the deployment of BETs across various U.S. ports and regions.

## 3. Methodology

A simulation-based analysis is performed in this study to assess the energy requirements of electric drayage trucks and understand the influence of seasonal and regional variations on the total energy consumption. This section introduces the proposed energy consumption model for simulating heavy-duty vehicles.

## 3.1 Truck Energy Consumption Model

The vehicles modeled in this study are representative of heavy-duty Class-8 diesel and electric trucks available in the U.S. market. Models for both vehicle types are developed in the MATLAB/Simulink environment. The block diagram of the vehicle simulator is depicted in Figure 3.

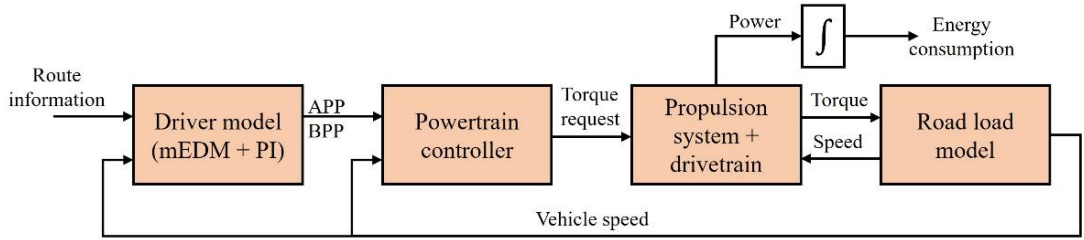

Figure 3: Block diagram of the developed vehicle simulator. The main input is the route information; the main output is the corresponding energy consumption (Onori, Serrao, and Rizzoni 2016).

The input to the simulator is the route information consisting of speed limit, road grade and other traffic related information, based on which the target vehicle speed is determined using a driver model. The parametric driver model used in this work is the modified Enhanced Driver Model (mEDM) which has been specifically developed for emulating the driving behavior of a human driver of a heavy-duty truck, including the distinct behavior of preemptively slowing down the vehicle before entering slow speed segments like highway-exits and junctions because of having high vehicle inertia (Shiledar et al. 2023). Additionally, the parametric approach followed in this driver model allows for emulating levels of driver aggressiveness (calm, conservative and aggressive) by changing the model calibration. However, in this work only the conservative driver aggressiveness is considered. The proportional-integrator (PI) controller in the driver model then actuates the accelerator (APP) and brake (BPP) pedals to minimize the error between the reference and actual vehicle speed. These driver commands are converted by the powertrain controller into torque requests for the propulsion system. The main propulsion component (i.e. engine in a diesel-truck and electric motor(s) in an electric truck) receives the torque request from the controller and generates the required torque. Finally, the power is delivered to the wheels through the drivetrain, to counter the resistive forces being applied on the vehicle, i.e., the road load. The next subsections provide the details about the approaches used for modeling the various components of the vehicle simulator.

### 3.1.1 Detailed Road Load Modeling

As far as resistive forces are concerned, only the ones impacting the longitudinal dynamics of the vehicle are included in the model, as illustrated in Figure 4.

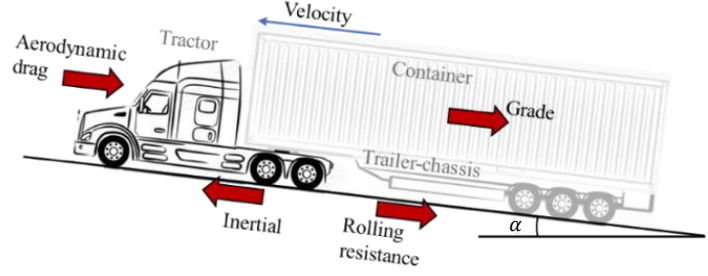

Figure 4: Longitudinal resistive forces on a truck. The vehicle's powertrain will need to generate the tractive or braking forces to counter the resistive forces and achieve a desired vehicle speed.

Aerodynamic drag, $F_{aero}$, rolling resistance, $F_{roll}$, inertial force, $F_{inertial}$ and grade force, $F_{grade}$ are the main resistive forces which add up in the road load, $F_{rl}$, that a truck must overcome:

$$F_{rl} = \underbrace{\frac{1}{2}\rho_a C_d(\phi,\psi) A_f v^2}_{F_{aero}} + \underbrace{MgC_{rr}(T_{tire},v)\cos(\alpha)}_{F_{roll}} + \underbrace{Mg\sin(\alpha)}_{F_{grade}} + \underbrace{M_{eq}\frac{dv}{dt}}_{F_{inertial}}$$
$$T_w = F_{rl} \cdot r_w$$
$$\omega_w = \frac{v}{r_w}$$
(1)

where aerodynamic drag depends on air density $\rho_a$, drag coefficient $C_d$, vehicle frontal area $A_f$, and speed $v$; rolling resistance depends on vehicle mass $M$, rolling resistance coefficient $C_{rr}$, and road slope $\alpha$ which is determined based on the elevation profile of the route; grade force depends on vehicle mass and road slope; inertial force depends on equivalent vehicle mass $M_{eq}$ and acceleration $\frac{dv}{dt}$; $g$ is the gravitational constant. $T_w$ is the total tractive wheel torque provided by the powertrain, $\omega_w$ is the wheel angular speed and $r_w$ is the wheel radius. It is thus critical for estimating energy consumption that these resistive forces are comprehensively modeled.

### *Aerodynamic Drag Coefficient Model*

A typical day of drayage operations involves picking up the shipping container at the port, delivering it to its intended destination, and then driving back to the port to perform the next set of deliveries. Depending on the leg of the operation, the vehicle parameters, specifically the weight (depending on cargo) and truck configuration (bobtail, tractor-trailer, or tractor-flatbed trailer), will differ. Utilizing the data from (Vegendla et al. 2016), the aerodynamic drag coefficient, $C_d(\phi,\psi)$, for the for conventional heavy-duty trucks, commonly used in the United States, is modeled based on various truck configurations, $\phi$, and as a function of the relative yaw angle, $\psi$, with the wind direction. The relative yaw angle with the wind is determined using the following equation from (Howell 2014),

$$\psi = \left| \tan^{-1}\left( \frac{v_w \sin(\theta_w - \theta_h)}{v + v_w \cos(\theta_w - \theta_h)} \right) \right| \tag{2}$$

where, $v_w$ is the wind speed. $\theta_w$ and $\theta_h$ are the wind and vehicle heading directions, respectively.

Figure 5 shows the yaw averaged aerodynamic drag coefficient for different configurations of conventional style trucks, along with the percent change in aerodynamic drag coefficient observed between the different truck configurations.

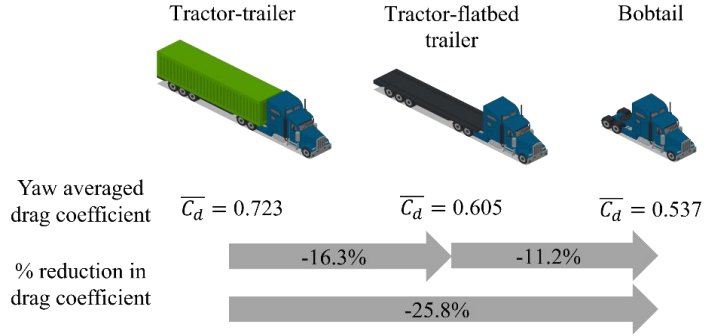

Figure 5: Percent reduction in yaw averaged (0- and 6-degree yaw angles) aerodynamic drag coefficient $C_d$ for different configuration of conventional style trucks. The drag coefficient can change up to 25.8% which will directly affect the vehicle's energy consumption.

## *Rolling Resistance Coefficient Model*

A detailed transient rolling resistance (RR) model specifically designed and calibrated for heavy-duty truck tires in (Hyttinen et al. 2022) is incorporated in the simulator. This model considers the variation in rolling resistance coefficient, $C_{rr}(T_{tire}, v)$, due to tire shoulder temperature, $T_{tire}$, and vehicle speed. The equations obtained for rolling resistance coefficient modelling are given below:

$$\begin{aligned} C_{rr}(T_{tire}, v) &= C_{rr,m}\big(T_{tire} + \Delta T(v)\big) \\ C_{rr,m}(T) &= C_{rr,h} + \big(C_{rr,0} - C_{rr,h}\big)e^{-\frac{T}{\lambda}} \\ \Delta T(v) &= T_h - (T_0 - T_h)\left(2 - \frac{2}{1 + e^{-\frac{v}{\alpha_v}}}\right) \end{aligned} \tag{3}$$

In the above model, the influence of vehicle speed, $v$ on RR coefficient is accounted for by a virtual shift in tire temperature given by $\Delta T(v)$ and the transient RR coefficient is given by master rolling resistance curve, $C_{rr,m}(T)$. Further, $C_{rr,0}$ represent the RR coefficient at 0 degC tire temperature, $C_{rr,h}$, indicate the RR coefficient at elevated temperatures, and $\lambda$, the decay parameter that determines the rate at which the RR value transitions from higher to lower values as the tire temperature changes. Furthermore, the tire temperature shift parameters include $T_0$ and $T_h$, which denote the temperature shift at zero and high vehicle speeds, respectively, while $\alpha_v$ governs the rate of temperature shift reduction from higher to lower values across varying speeds.

Additionally, a simplified thermal dynamics model for modeling the heating/cooling effects due to the interaction between tire, road surface, and ambient air, $T_{amb}$, is adopted from (Hyttinen et al. 2023). The thermal dynamics model of tire heating/cooling is given as follows,

$$\frac{dT_{tire}}{dt} = -\frac{1}{\tau(v)}\big(T_{tire} - T_{st}(v, T_{amb})\big)$$
$$\tau(v) = \tau_h + (\tau_0 - \tau_h)e^{-\frac{v}{\delta_\tau}} \tag{4}$$
$$T_{st}(v, T_{amb}) = kv + n - (n - T_{a,ref})e^{-\frac{v}{\gamma}} - (T_{a,ref} - T_{amb})$$

where, $T_{st}(v, T_{amb})$ is the stabilized tire temperature for a given vehicle speed and ambient temperature which depends on calibrating multiple factor that include $k, n$ and $\gamma$. $T_{a,ref}$ is the reference ambient temperature where the model is parameterized. Furthermore, $\tau(v)$ is the vehicle speed-dependent thermal time constant. The thermal constant varies between $\tau_0$ for when the vehicle is stationary and $\tau_h$ for higher vehicle speeds. $\delta_\tau$ govern the rate of decline from high to low value of the time constant at different speeds.

### 3.1.2 Powertrain Modeling

Figure 6 shows the schematic of the powertrain along with the auxiliary components for the conventional (diesel-powered) and the electric truck. The details of conventional and electric truck powertrains are provided next.

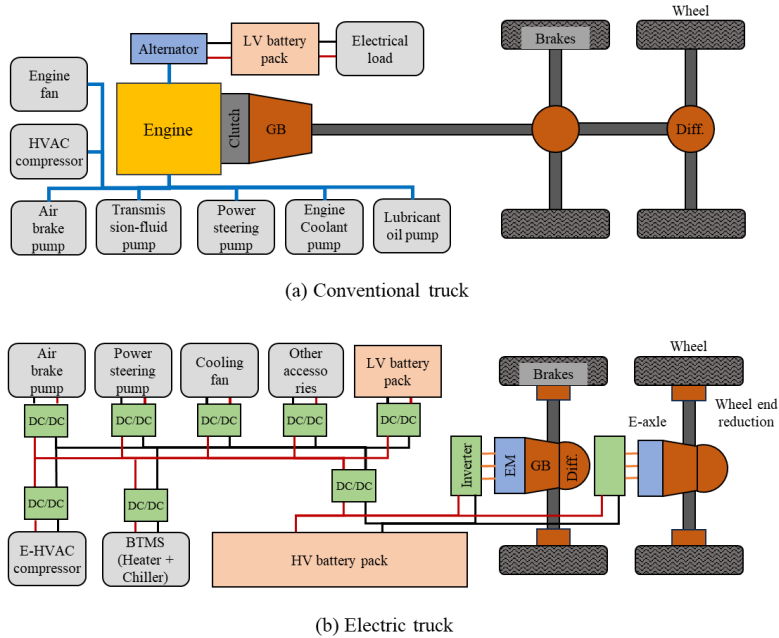

Figure 6: Schematic of (a) conventional (diesel) and (b) battery electric truck powertrains along with auxiliary components. GB: gear box; Diff.: differential gear; DC/DC: converters; LV: low voltage; EM: electric motor.

## Conventional Truck Powertrain Model

The powertrain of the conventional truck (Figure 6(a)) consists of a 339.2 kW (455hp) diesel powered engine from the 2018 model year with a 10-speed automatic manual transmission and a final rear drive reduction of 2.64:1 at the differential. The engine fuel consumption map as shown in Figure 7(a) is obtained from the EPA Greenhouse Gas Emissions Models (GEM) vehicle simulator, while the transmission gear ratios and efficiencies are obtained from the Eaton Fuller 10-speed transmission specification sheet. The truck specifications have been selected to be representative of real trucks used in drayage operations, according to the report in (Papson and Ippoliti 2013). Table 1 presents how the performance in terms of maximum speed, gradability, and startability of the truck model compares to the requirements reported in (Papson and Ippoliti 2013), showing good agreement. The model equations for the conventional truck powertrain are given below.

$$T_{ax} + T_{br} = T_w$$

$$T_{eng} = \frac{T_{ax}}{\lambda_{gb,i} \cdot \lambda_{fd}} \eta_{gb,i}^\gamma \cdot \eta_{fd}^\gamma + T_{aux}(T_{amb}) \begin{cases} \gamma = 1, & T_{ax} < 0 \\ \gamma = -1, & T_{ax} \geq 0 \end{cases} \quad (5)$$

$$\omega_{eng} = \omega_w \cdot \lambda_{gb,i} \cdot \lambda_{fd}$$

$$\dot{m}_f = \dot{m}_f(\omega_{eng}, T_{eng})$$

where $T_{ax}$ is the total axle torque and $T_{br}$ is the total brake torque. $T_{eng}$ and $\omega_{eng}$ are the engine torque and angular speed, respectively. $\lambda_{gb,i}$ and $\eta_{gb,i}$ is the gear reduction and gear efficiency of $i$-th gear in the gearbox, respectively, while $\lambda_{fd}$ and $\eta_{fd}$ are the gear reduction and efficiency of the final drive reduction, respectively. $T_{aux}(T_{amb})$ is the additional torque required to power the auxiliaries. Finally, $\dot{m}_f$ is the fuel consumption rate of the engine. The details of modeling the auxiliary components of the conventional truck will be explained in a later section.

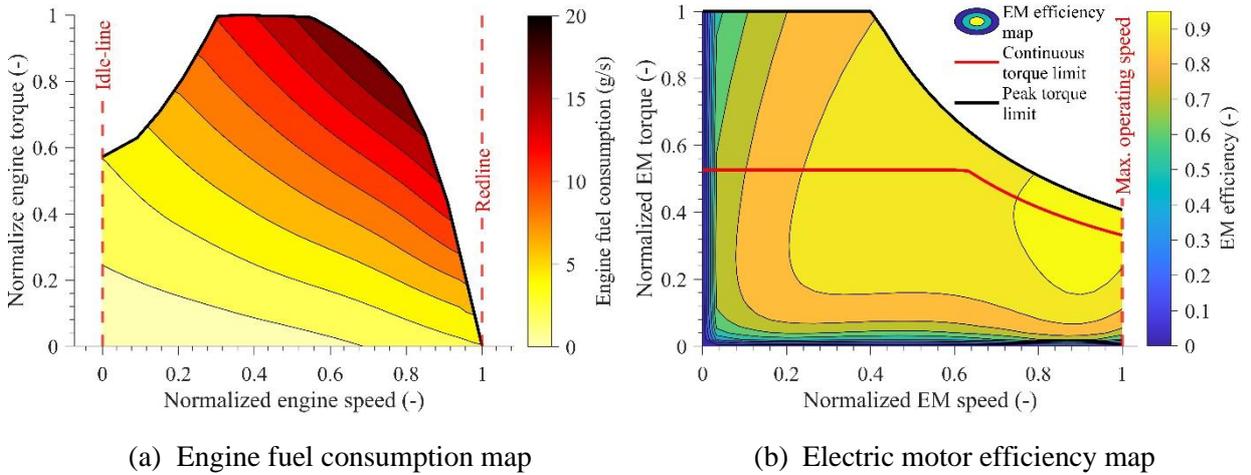

(a) Engine fuel consumption map  (b) Electric motor efficiency map

Figure 7: Normalized performance maps: (a) Fuel consumption map of the 455-hp engine from model year 2018; (b) Power conversion efficiency map of 250 kW electric motor in the e-axle.

## *Electric Truck Powertrain Model*

The powertrain of the electric truck (Figure 6(b)) is a tandem e-axle configuration based on the information publicly available on the Kenworth 6x4 T680E ("T680E | Kenworth," n.d.). Each e-axle has an electric motor (EM) that can provide a maximum power of 250 kW, and their power conversion efficiency is modeled using a performance map as illustrated in Figure 7(b). The EM is connected to a 3-speed gearbox (with gear ratios of 5.6:1, 2.8:1, and 1.4:1). The gearbox along with the final drive ratio of 2.47:1 and wheel end reduction of 2:1 give the required torque amplification at the wheels to have a similar or better performance for the electric truck over the conventional baseline in terms of maximum speed, acceleration times, gradability and stability specification requirements for a heavy-duty truck in drayage applications, as shown in Table 1. The comparison shows that the electric truck modeled in this study has performance capabilities which match or exceed the requirements for drayage application. The model equations for the electric truck powertrain are given below.

$$T_{ax} + T_{br} = T_w$$

$$T_{em} = \frac{T_{ax}}{n_e \cdot \lambda_{gb,i} \cdot \lambda_{fd} \cdot \lambda_{we}} \eta_{gb,i}^\gamma \cdot \eta_{fd}^\gamma \cdot \eta_{we}^\gamma \quad \begin{cases} \gamma = 1, & T_{ax} < 0 \\ \gamma = -1, & T_{ax} \geq 0 \end{cases}$$

$$\omega_{em} = \omega_w \cdot \lambda_{gb,i} \cdot \lambda_{fd} \cdot \lambda_{we} \tag{6}$$

$$P_{em} = T_{em} \cdot \omega_{em} \cdot \eta_{em}^\gamma(\omega_{em}, T_{em}) \quad \begin{cases} \gamma = 1, & T_{em} < 0 \\ \gamma = -1, & T_{em} \geq 0 \end{cases}$$

$$I_{batt} = \frac{n_e \cdot P_{em} + P_{aux}(T_{amb})}{V_{batt}}$$

where $T_{em}$, $\omega_{em}$, $P_{em}$ and $\eta_{em}$ are the EM torque, angular speed, electrical power and efficiency, respectively. $n_e (= 2)$ is the number of e-axles, with $\lambda_{we}$ and $\eta_{we}$ being the gear reduction and efficiency of the wheel end reduction. $I_{batt}$ and $V_{batt}$ are the current and terminal voltage of the high-voltage Li-ion battery pack, respectively. $P_{aux}(T_{amb})$ is the total power drawn from the battery pack due to all the electrical auxiliaries. The details of the electric truck auxiliary power components will be discussed in a later section.

Table 1: Specification and performance comparison of conventional and electric truck

| | Powertrain | Drayage application requirements | Conventional truck | Electric truck |
|---|---|---|---|---|
| **Specification** | Engine/e-axle power (kW) | ≥295 | 339.2 | 500 (250 kW for single E-axle) |
| | Transmission/gearbox | No specific requirements | 10 speeds | 3 speeds |
| | Final drive ratio | | 2.64 | 2.47 |
| | Wheel-end reduction ratio | | NA | 2 |
| **Performance** | Maximum speed (km/h) | ≥100 | 129.5 | 141.9 |
| | Acceleration 0-48 km/h time (s) | Must be sufficient to operate on local roads and highways | 13 | 12 |
| | Acceleration 48-80 km/h time (s) | | 23 | 21 |
| | Acceleration 80-96 km/h time (s) | | 24 | 17 |
| | Gradeability 1% (km/h) | No specific requirements | 121.9 | 135.7 |
| | Gradeability 2% (km/h) | | 90.8 | 118.1 |
| | Gradeability 6% (km/h) | 64 | 60.2 | 67.6 |
| | Startability (%) [EM continuous torque limit] | ≥6 | 35 | 17 |
| | Startability (%) [EM peak torque limit] | ≥6 | 35 | 34 |

➤ *Gradeability* is defined as the maximum speed the vehicle can travel at a constant grade.
➤ *Startability* is defined as the maximum grade at which the vehicle can start from being stationary and can sustain a minimum speed of 8 kmph (5 MPH).
➤ All performance capabilities are determined for vehicle weight of 27.2 Ton (60,000 lbs.) at 25 °C ambient temperature.
➤ Electric truck acceleration and gradability are determined using EM continuous torque limits.

To estimate the energy usage of the vehicle, a high voltage (HV) Li-ion battery pack model is integrated into the vehicle simulator, capable of dynamically responding to the power demands of the drivetrain. The battery pack construction is based on an LG INR21700-M50T cell which uses NMC-811 material in the cathode and graphite/silicone in the anode (Pozzato, Allam, and Onori 2022). The battery model captures both electric and thermal dynamics of the cells. For simplicity, the assumption of homogeneity, i.e., that all cells in the pack are identical, is considered here for modeling the HV battery pack. This allows for direct scaling of the inputs/outputs of a single cell to pack-level inputs/outputs. Neglecting cell-to-cell variations in a battery pack with parallel-connected cells was shown in (Lucero, Sujan, and Onori 2024) to insignificantly affect the energy output of the pack. The electrical dynamics of an individual cell is modeled using a first-order equivalent circuit model (ECM) (Plett 2015), as shown in Figure 8.

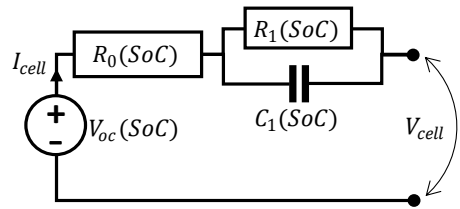

Figure 8: First-order ECM for modeling the electrical dynamics of the cell.

$$\frac{dV_{RC}}{dt} = -\frac{1}{\tau(SoC)}V_{RC} + \frac{1}{C_1(SoC)}I_{cell} \qquad (7)$$

$$V_{cell}(t) = V_{oc}(SoC) - R_0(SoC) \cdot I_{cell} - V_{RC}(t) \qquad (8)$$

where the polarization voltage is $V_{RC}$, the capacitance is $C_1$, the relaxation time scale is $\tau(SoC) = R_1(SoC) \cdot C_1(SoC)$, the polarization resistance is $R_1(SoC)$, the high-frequency resistance is $R_0$, and the cell current[1] is $I_{cell}$. The model parameters depend on the cell state-of-charge (SoC) which is obtained by Coulomb counting:

$$SoC(t) = SoC(t_0) - \frac{1}{3600 \cdot Q_{nom}}\int_{t_0}^{t} I_{cell}(s)\, ds \qquad (9)$$

The lumped-element thermal model of the cell is similar to the one used in (Lin et al. 2013) where assuming fast equilibration between the cell's core and surface temperatures results in a single-state model:

$$C_{cell}\frac{dT_{cell}}{dt} = \underbrace{I_{cell} \cdot \big(V_{oc}(SoC) - V_{cell}(t)\big)}_{\dot{Q}_{loss,cell}} + \frac{1}{R_{th}}\big(T_{amb} - T_{cell}(t)\big) \qquad (10)$$

By assumption, we take the parameters of the thermal model $C_{cell}$ and $R_{th}$ to be independent of SoC.

To calibrate the electric and thermal sub models, the procedure described in (Lucero, Sujan, Onori 2024) is followed. Experimental data from a Hybrid Pulse Power Characterization (HPPC) test, done for a fresh battery cell, is used (Pozzato, Allam, and Onori 2022). After calibration, the model is validated experimentally by using a repeated urban dynamometer driving schedule (UDDS)-based drive cycle test where a root-mean-square error of 1.10 mV and 0.005 °C errors between the experimental data and the model for the voltage and temperature, respectively is achieved as seen from Figure 9.

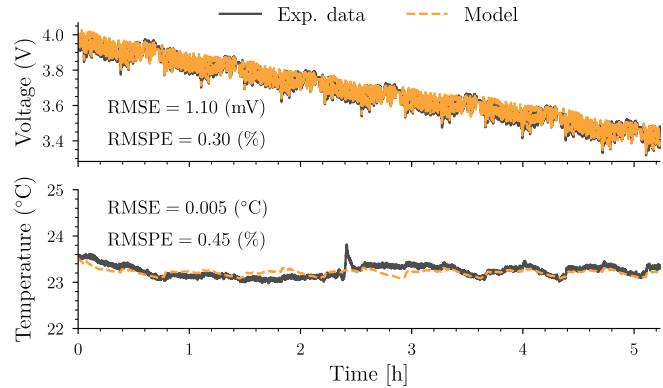

Figure 9: Battery cell model valication using experimental UDDS driving cycle data.

---

[1] Here the following sign convention is followed such that the positive (negative) cell current $I_{cell} > 0$ ($I_{cell} < 0$) denotes discharge (charge).

Like the cell, the battery pack's temperature dynamics are modeled using a lumped-element model. The schematic of the battery pack housing lumped-element thermal dynamics model is shown in Figure 10. The battery pack housing temperature $T_h$ follows the dynamics as given by the following equation which is obtained from (Teichert, Schneider, and Lienkamp 2022).

$$C_h \frac{d}{dt} T_h = N_{cells} \cdot \frac{T_{cell}(t) - T_h(t)}{R_{cell,h}} + \frac{T_{amb}(t) - T_h(t)}{R_{h,amb}} + \dot{Q}_{cool} \cdot u_c + \dot{Q}_{heat} \cdot u_h \qquad (11)$$

The first term represents the heat contributions of the total number of cells to the housing temperature mediated by the thermal resistance $R_{cell,h}$. The second term models the exchange of heat between the battery pack housing and the ambient environment mediated by the thermal resistance $R_{h,amb}$. The third and fourth terms represent the heat/cooling rejection capabilities of the battery thermal management system. The variables $u_c, u_h \in \{0,1\}$ are binary variables that controls whether the thermal management system is cooling or heating the pack housing or is otherwise off.

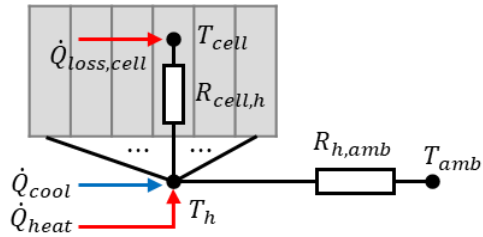

Figure 10: Schematic of lumped-element thermal dynamics model of the battery pack.

The temperature dynamics of the individual cell, described in equation (10), are then modified to have the pack housing act as its thermal surroundings:

$$C_{cell} \frac{dT_{cell}}{dt} = I_{cell} \cdot (V_{oc}(SoC) - V_{cell}(t)) + \frac{1}{R_{th}} (T_h(t) - T_{cell}(t)) \qquad (12)$$

where the first term on the right-hand side represents the heating losses in a cell, while the second term is the heat exchanged between the cell and the pack housing, which earlier in equation 10 was the heat exchange between the cell and the surrounding air. Since all the cells are assumed to be in thermal equilibrium, there is no term that represents the heat transfer between adjacent cells of the pack.

*Auxiliary Components Modeling*

Operation of auxiliary components (heater and chiller of the battery thermal management system (BTMS), power steering system, transmission fluid pump, pneumatic brake system, etc.) is critical for the functioning of an HDV. In addition, the weather conditions affect the need to heat or cool the cabin for the driver's comfort through the HVAC system. The energy required for powering the auxiliaries in an HDV can be substantial. Given that in an electric powertrain all the energy is eventually drawn from the batteries, it is extremely critical to model the power consumption of the auxiliary components to accurately estimate the vehicle's overall energy consumption. However, dynamic modeling of all the auxiliary loads of an HD truck would be very complex as their operation depends on driver preference, road and weather conditions, and traffic circumstances (Andersson 2004). Since only the power consumption of the auxiliaries is of interest, the dynamic models of different auxiliary systems like in (Khuntia et al. 2022) for cabin-HVAC or (Silvas et al. 2013) for power steering pump, which would be suitable for control-oriented applications, would be much more complicated than necessary. Thus, the simpler transient auxiliary-load consumption model from (Gao et al. 2014) is adopted for modeling various auxiliary components of the electric truck in this work. The auxiliary loads are modeled using an on/off duty-cycle-based approach, where each auxiliary component is periodically turned on and off during the vehicle operation. The auxiliary component time-period, duty cycle, and operating power when turned on for both the conventional and electric trucks are obtained from (Gao et al. 2014) and are tabulated in Tables 2 and 3, respectively.

Table 2: Duty-cycle based modelling of conventional truck auxiliary component's power consumption

| Auxiliary component | Period (s) | Duty (%) | Operating power as a function of engine speed (kW) | | | | |
|---|---|---|---|---|---|---|---|
| Engine fan | 200 | 5 | 0.71 | 2.36 | 7.56 | 16.73 | 29.81 |
| HVAC compressor | 150 | Amb. TD | 1.03 | 2.45 | 3.83 | 5.27 | 6.67 |
| Air-brake pump | 100 | Amb. TD | 1.0 | 1.69 | 2.53 | 3.51 | 4.51 |
| Power steering pump | 100 | 10 | 3.04 | 5.52 | 8.04 | 10.47 | 12.99 |
| Engine coolant pump | - | - | 0.0 | 0.22 | 0.56 | 1.25 | 1.88 |
| Lubricant oil pump | - | - | 0.14 | 1.24 | 2.05 | 3.24 | 4.97 |
| Transmission fluid pump | - | - | 0.5 | | | | |
| Electrical load | - | - | 0.6 | | | | |
| **Engine speed (RPM)** | | | 500 | 1000 | 1500 | 2000 | 2500 |

➢ Amb. TD = Ambient temperature dependent

Table 3: Duty-cycle based modelling approach of electric truck auxiliary component's power consumption

| Auxiliary component | Period (s) | Duty (%) | Operating power (kW) |
|---|---|---|---|
| BTMS | - | - | Battery TD |
| E-HVAC compressor | 150 | Amb. TD | 2.71 |
| Air-brake pump | 100 | Amb. TD | 2.19 |
| Power steering pump | 100 | 10 | 3.0 |
| Cooling fan | 200 | 4 | 5.49 |
| EM cooling pump | - | - | 0.16 |
| Transmission fluid pump | - | - | 0.5 |
| Electrical load | - | - | 0.5 |

➢ Amb. TD = Ambient temperature dependent
➢ Battery TD = Battery temperature dependent

Most of the auxiliary components in a conventional truck are powered by the direct mechanical coupling with the engine. Thus, their power consumption is sensitive to the engine speed. Conversely, in an electric truck, all the auxiliaries are independent of vehicle speed and are powered individually, therefore they can be operated at the most efficient point.

Ambient temperature is known to influence certain auxiliary loads. To take this into account heating and cooling loads on the HVAC system and the air brake system load are considered as a function of ambient temperature, as illustrated in Figure 11. The relationships used herein are established based on the information obtained on the steady-state load on these auxiliary systems under varying temperature conditions (Okaeme et al. 2021; Li et al. 2021; Sjostedt, Wikander, and Flemmer 2014). The duty cycle of these components is altered from the nominal operation duty cycle of 50% and 5% for the HVAC compressor and air-brake pump, respectively. This change is performed based on the ambient temperature-dependent load such that the time-averaged power of the component based on the modified duty cycle, matches with the corresponding ambient temperature-dependent load.

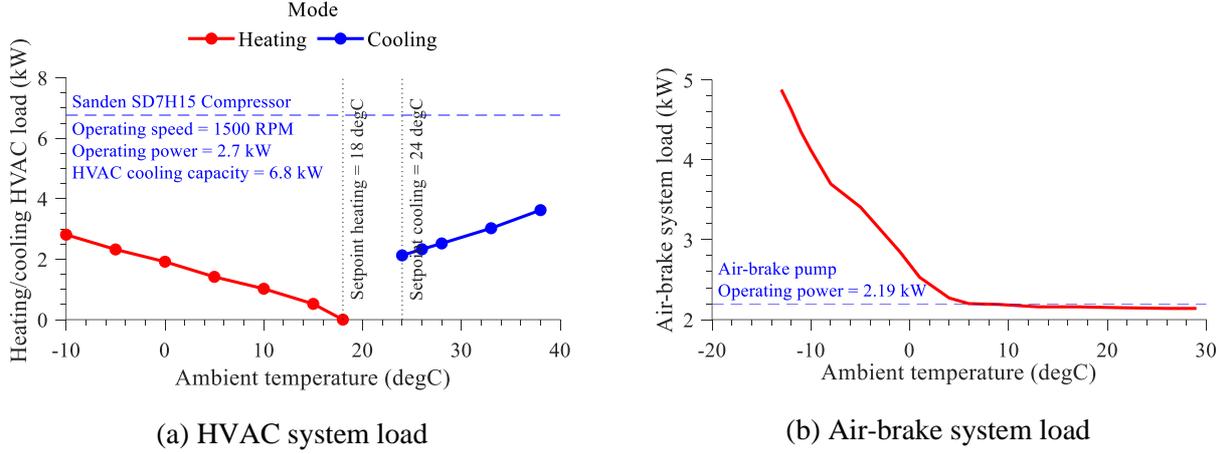

(a) HVAC system load  (b) Air-brake system load

Figure 11: Ambient temperature dependent (a) HVAC system and (b) air-brake system loads.

Finally, for the electric truck, the BTMS is modeled separately, considering the thermal model of the battery pack and a simple heuristic based thermal management strategy developed to maintain the battery temperature within a $\pm 5°C$ about the desired operating temperature of 25°C. The power consumption of BTMS heater, $P_{heater}$, and chiller, $P_{chiller}$, are determined based on the constant value of heat addition and extraction from the housing as well as on the coefficient of performance, $COP$, of the BTMS system for heating ($COP_{BTMS,heater} = 4$) and cooling ($COP_{BTMS,cooler} = 3$) (Teichert, Schneider, and Lienkamp 2022) using the equation below:

$$P_{heater} = \frac{\dot{Q}_{heat}}{COP_{BTMS,heater}}, \quad P_{chiller} = \frac{\dot{Q}_{cool}}{COP_{BTMS,chiller}} \tag{13}$$

## *3.2 Truck Energy Consumption Model Validation*

To validate the energy consumption model developed in the previous section, data from a recent study that investigated the effects of ambient conditions on fuel consumption in commercial vehicles was utilized (Surcel 2022). This study examined the variation in fuel consumption due to seasonal changes in Quebec, Canada. The track test results of heavy-duty 2017 Freightliner Cascadia trucks from this study were used to validate the energy consumption model of the conventional truck. The specifications of the Freightliner Cascadia trucks in the referenced study and the conventional truck modeled in this work are similar in many aspects, with the primary differences being the vehicle model years (2017 vs. 2018) and engine power ratings (410 hp vs. 455 hp).

The tests conducted during the fall and winter seasons in the study were recreated in simulations. The vehicle used in the tests had a weight of 24,880 kg and was driven at a constant speed on oval-shaped tracks. The vehicle's average speed, test duration, ambient conditions, and the corresponding fuel consumption—both from experimental measurements and simulations—are presented in Table 4. Additionally, the speed and fuel consumption profiles for the simulated tests are illustrated in Figure 12.

Table 4: Test conditions and vehicle energy consumption model validation for conventional truck

| Season | Fall | | Winter | |
|---|---|---|---|---|
| Average vehicle speed (kmph) | 105 | | 90 | |
| Test run length (km) | 100 | | 90 | |
| Average ambient temperature (degC) | 15.4 | | -6.3 | |
| Average air density (kg/m$^3$) | 1.245 | | 1.337 | |
| **Average fuel use** | **Experiment** | **Simulation** | **Experiment** | **Simulation** |
| Consumed fuel (kg) | 29.45 | 29.65 | 27.76 | 26.98 |
| Fuel consumption (L/100 km) | 35.27 | 35.51 | 36.94 | 35.90 |
| Fuel consumption increase between fall to winter season (%) | (-) | (-) | 4.7% | 1.1% |
| Fuel consumption difference between experiment and simulation (%) | -0.7% | | 2.8% | |

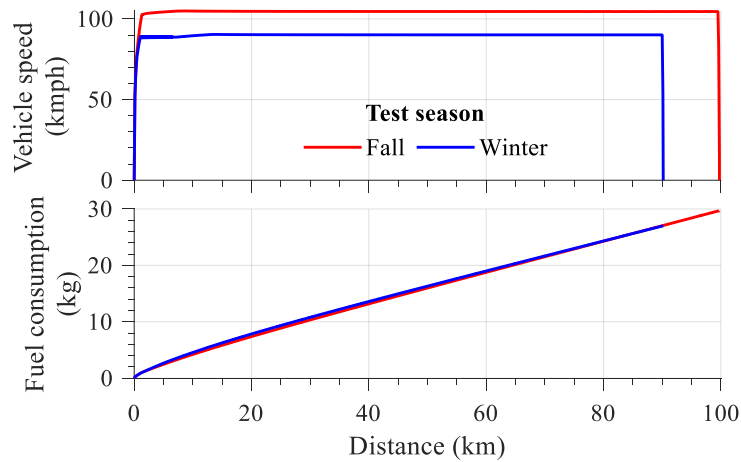

Figure 12: Vehicle speed and fuel consumption for simulated tests in fall and winter season.

The fuel consumption values obtained from simulations closely match the experimental data, with a deviation of -0.7% in the fall season and 2.8% in the winter season. This strong agreement validates the accuracy of the proposed energy consumption model for the heavy-duty conventional truck. However, due to the limited availability of experimental data for electric trucks, a system-level validation of the energy consumption model for electric trucks is currently not feasible. Nonetheless, it is important to highlight that the modeling of various components in the electric truck powertrain involved the use of component-level experimental data. Moreover, the energy consumption estimates presented in the following sections, based on the proposed electric truck model, are within the range of 120-180 kWh/100 km, as suggested in the reviewed literature and preliminary electric truck testing ("Electric Trucks Have Arrived: The Use Case for Heavy-Duty Regional Haul Tractors" 2022; "Battery Electric Truck and Bus Energy Efficiency Compared to Conventional Diesel Vehicles" 2018).

## *3.3 Routes and Weather Data Collection*

To determine the impact of seasonal and geographic variations on the energy consumption of drayage trucks, the first data acquired are the route origins and destinations (OD) that these trucks are expected to travel on in a year in a specific region. The route ODs are obtained using the process described in (Sujan et al., 2024), which takes into account the freight movement information collected using the Streetlight data. For the analysis presented in this study, only the route ODs obtained for the drayage trucks operating in the Port of Savannah, GA, are considered. The route ODs are identified separately by month and for both the inbound and outbound travel directions to/from the Port of Savannah. Figure 13 shows the major origin and destination locations for inbound and outbound travel to/from the port, in different months of the year 2021, respectively.

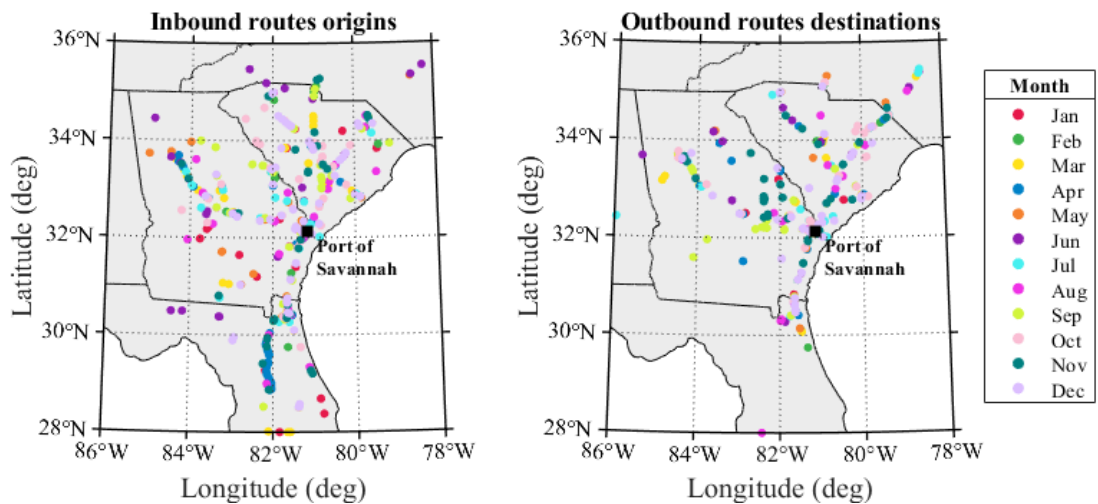

Figure 13: Locations of monthly inbound routes origins (left) and outbound routes destinations (right) for drayage routes at the Port of Savannah, GA.

Next, the route information of all the inbound and outbound trips for each of the OD locations, to and from the Port of Savannah, respectively, are determined using Google Routing and HERE Routing. The route information obtained consists of the latitude and longitude coordinates, distance, road speed limits, and elevation (road grade) over the route. After the route OD identification and route information extraction, the next step involves the acquisition of local weather conditions along those routes. The impact of weather on overall vehicle energy consumption is substantial, with ambient temperature and air density playing key roles, as shown in equations 1 – 8. Leveraging the meteorological NOAA MADIS dataset and applying the algorithm developed in (Moore, Siekmann, and Sujan 2023a) the spatial-temporal variations in ambient temperature and air density at each GPS coordinate along a given route are obtained. For the purposes of this study, the weather dataset from 2021 is used to maintain consistency with the route data. The weather data for each route is specifically extracted for three distinct dates within the month that the route corresponds to. These selected dates represent typical driving scenarios, encompassing conditions on a hot day, a cold day, and a day with nominal temperatures. Collecting the weather data for these three specific days ensures that the maximum variability of weather conditions along the routes is captured. Further, the sampling of weather conditions on the route is conducted at a specific time of the day only, which in this case is selected to be noon, as it coincides with peak time of travel. Doing so, the second-by-second temporal changes in weather conditions over the route are lost, but it significantly reduces the computational burden during the simulation runtime. The process for identifying the dates to obtain the data is provided in the Appendix A.

The weather collection dates for all the routes are determined and the ambient temperature and air density data are obtained for all the routes. Figure 14 shows the monthly variation in average route ambient temperature and air density on cold, nominal, and hot days, obtained for all the routes in the region around the port of interest. In addition to this, the bottom box plot of the figure shows the variation of day type by temperature in each month. The climate of this region is a humid subtropical climate with short, mild winters and long, hot summers. Still, depending on the route being traveled in this region and the time of year, the weather conditions can vary significantly. Comparing the coldest days in winter months (November-February) with the warmest days in summer months (June-August), the air density differs by roughly 10%. Similarly, the average route temperature can go as low as 5°C to a high value of nearly 35°C. These variations in weather conditions can significantly alter the road load, specifically the aerodynamic drag which depends directly on the air density as well as the rolling resistance which is indirectly influenced by the ambient temperature. As seen previously, some of the auxiliaries' power consumption is ambient temperature dependent and these weather variations can also profoundly influence the total auxiliary power consumption.

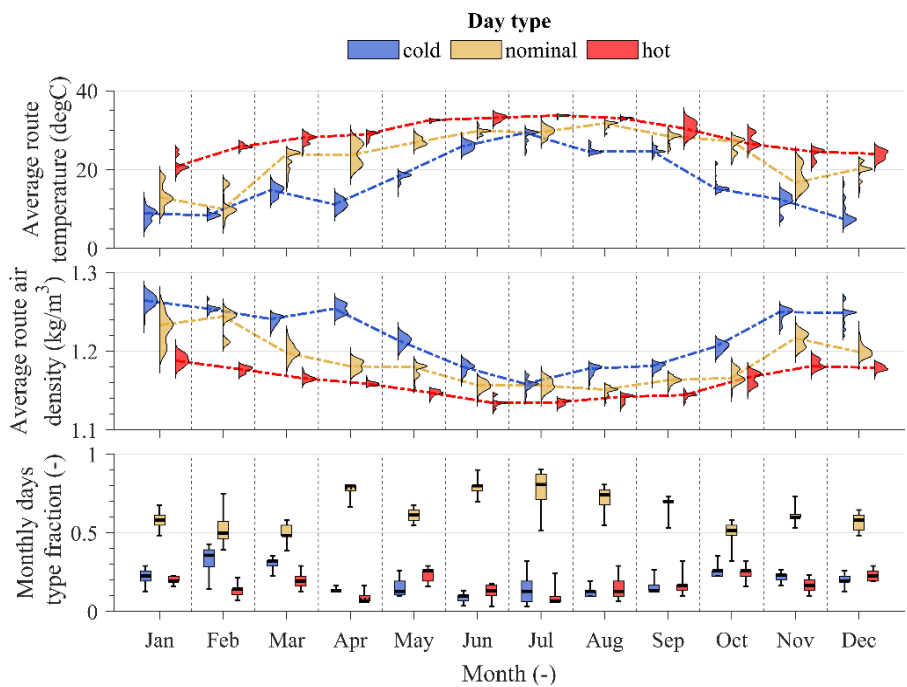

Figure 14: Monthly weather condition variations in the region around the Port of Savannah, GA. Each month is characterized with average route temperature (top) and average route air density (middle) for cold, nominal, and hot

representative days. The distributions show the variations over the various routes. The box plots (bottom) show the fraction of days that falls within each day type for each month.

A large-scale simulation study is conducted using these weather conditions on different routes as input variables to the vehicle models developed earlier, to understand the impact of varying weather conditions on the energy consumption in the vehicle and infer any seasonal or geographical trends present in the energy consumption results.

## *3.4 Parametric Sensitivity Analysis on Energy Consumption*

Before analyzing the results of the large-scale simulation study, it is essential to understand the importance of geographical and weather conditions in assessing energy consumption. To illustrate this, a parametric sensitivity analysis is conducted on energy consumption, considering key factors, which include vehicle parameters (such as vehicle mass, aerodynamic drag coefficient, frontal area, rolling resistance coefficient, and accessory loads), geographical factors (including road speed and grade), and weather conditions (such as ambient temperature and air density). This sensitivity analysis examines each critical parameter by applying a ±10% variation one at a time.

To establish a baseline, 200 randomly selected routes from the Port of Savannah, GA region, along with associated weather conditions, are used, with a 27.2 Ton (65,000 lbs.) tractor-trailer configuration as the reference vehicle. Baseline energy consumption for these routes is calculated by simulating both electric and conventional trucks. Subsequently, energy consumption is recalculated with ±10% variations in each critical parameter. Using these simulation results, the sensitivity of model parameters is determined for battery energy consumption in electric trucks and fuel consumption in conventional trucks, as illustrated in Figures 15 and 16, respectively.

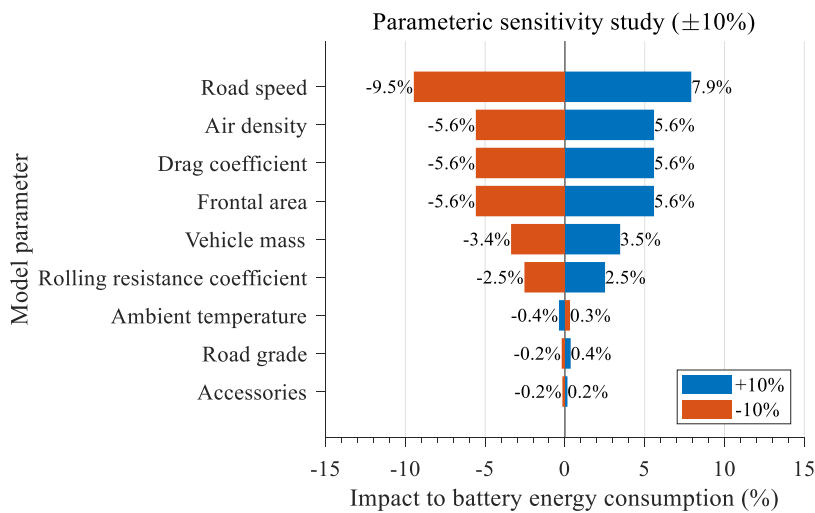

Figure 15: Parametric sensitivity of key factors (±10% change) on battery energy consumption for electric drayage trucks.

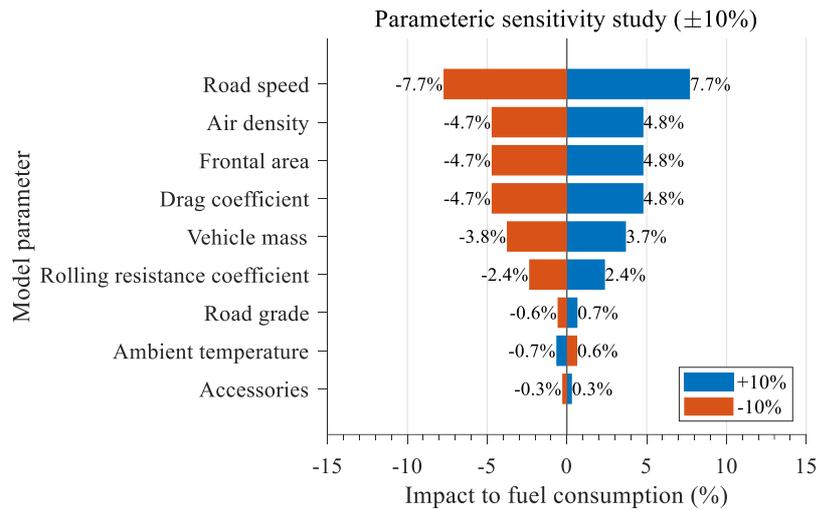

Figure 16: Parametric sensitivity of key factors (±10% change) on fuel consumption for conventional drayage trucks.

As anticipated, the road speed limit, which influences the vehicle speed, is the most sensitive factor to energy consumption in both electric and conventional drayage trucks. This is followed by factors such as air density, frontal area and drag coefficient, which directly affect the aerodynamic drag resistive force acting on the truck. Finally, vehicle mass and rolling resistance coefficient are also key sensitive factors. The sensitivity analysis performed here aligns with the findings presented in (Moore, Siekmann, and Sujan 2023b). This sensitivity analysis establishes the importance of accurately accounting these critical factors in the energy consumption analysis, which is primarily the aim of this work.

# 4. Results

In this section, the results of the simulation-based study are presented. The large-scale simulations required for this regional-level analysis are conducted using the high-performance computing resources of the Ohio Supercomputer Center (Ohio Supercomputer Center 1987). Access to multiple CPU cores on the supercomputer enables the simultaneous execution of multiple simulations with different combinations of routes, weather conditions, and vehicle weights, which significantly reduces the total time required to complete the simulations.

## *4.1 Total Energy Consumption Variations due to Geographical and Seasonal Variations at Port of Savannah, GA*

Firstly, the inbound and outbound routes from Figure 13 are grouped together to form different groups. The inbound and outbound route are clustered into 15 groups as shown in Figure 17.

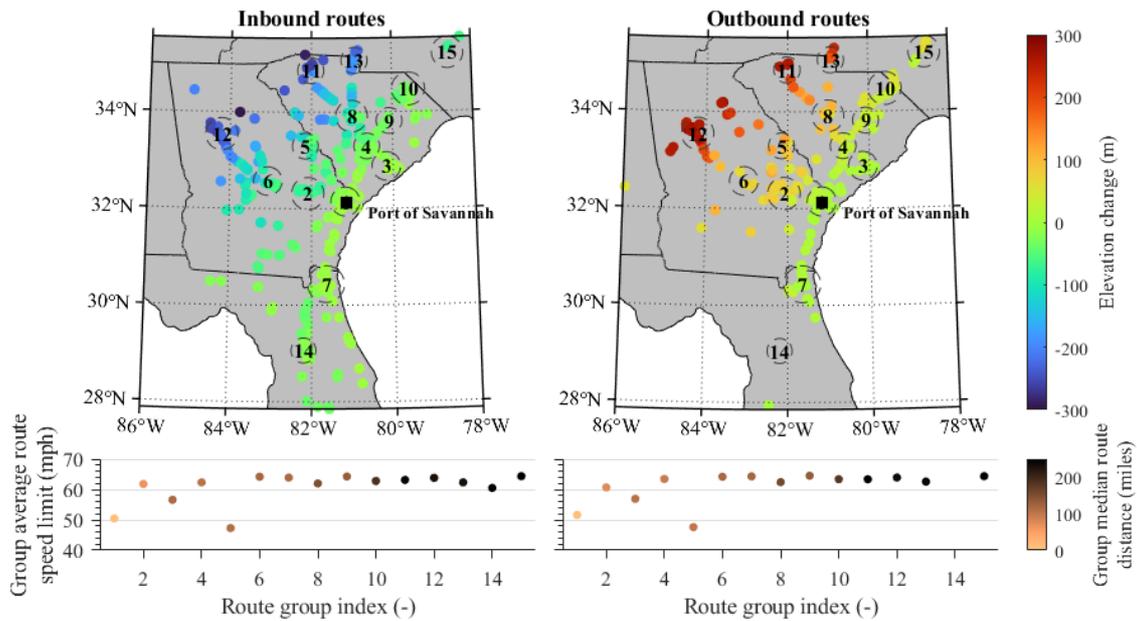

Figure 17: Inbound (left) and outbound (route) route grouping for analyzing the impact of geographical and weather variation on energy consumption in HD trucks at Port of Savannah, GA. The maps show the location of origins and destinations with the color scale indicating the elevation change. The charts at the bottom show the group average route speed limit with the color scale indicating the median route distance.

Since not all routes are present in every month, grouping the routes allows for analyzing the influence of weather variations on energy consumption to travel to (from) different locations around the Port of Savannah, throughout the year. The groups are numbered in ascending order based on the group center distance from the port. Additionally, the marker colors in the maps represent the observed elevation changes relative to the port. The peak elevation change is around ±300 m in the region around the port among all the routes. The group average speed limit is also shown in the same figure.

Investigating the energy consumption of travel to and from different route groups to the port enables an in-depth study of how geographical route conditions affect energy usage. Specifically, groups 1, 5, 6, 7, and 12 are of particular interest for both inbound and outbound routes. Routes in group 1 are the shortest, have negligible elevation change, and feature the lowest average speed limit among all the route groups. Groups 5, 6, and 7 have routes of similar distances but differ in elevation changes and average speed limits. For instance, groups 5 and 6 have similar elevation changes; however, the average speed limit for routes in group 5 is approximately 48 mph, which significantly lower than the 65 mph average speed limit for routes in group 6. Conversely, groups 6 and 7 have similar average speed limits, but group 6 routes experience significant elevation changes while group 7 routes have practically none. Finally, group 12 consists of the longest routes with the maximum elevation change. Comparing the energy consumption across these routes aids in understanding the impacts of geographical and seasonal factors.

The influence of geographical and seasonal variations on the energy consumption of a drayage truck over a year is illustrated in Figures 18 and 19. Figure 18 shows the fuel consumption in L/100 km for inbound and outbound routes with a conventional loaded truck weighing 27.2 Ton (60,000 lbs). The variation in energy consumption throughout the year results from differences in weather conditions, which directly affect road load, particularly aerodynamic drag and rolling resistance, and indirectly through auxiliary power consumption. For longer routes (groups 5, 6, 7, and 12), fuel consumption varies on average by -4% to 8% from the median due to seasonal variations. In contrast, shorter routes (group 1) exhibit higher variations, ranging from -12% to 16%. Similar patterns are observed for electric truck electric energy consumption (kWh/100 km), as shown in Figure 19. Electric energy consumption on longer routes varies by -3% to 11% from the median, while shorter routes show a variation of -16% to 25%. Additionally, energy consumption is slightly higher on outbound routes compared to inbound routes. This can be attributed to the relative change in elevation between the start and end points of the respective trips. Since the port is located at sea level, inbound trips in general experience a net negative elevation change, while outbound trips encounter a positive elevation change.

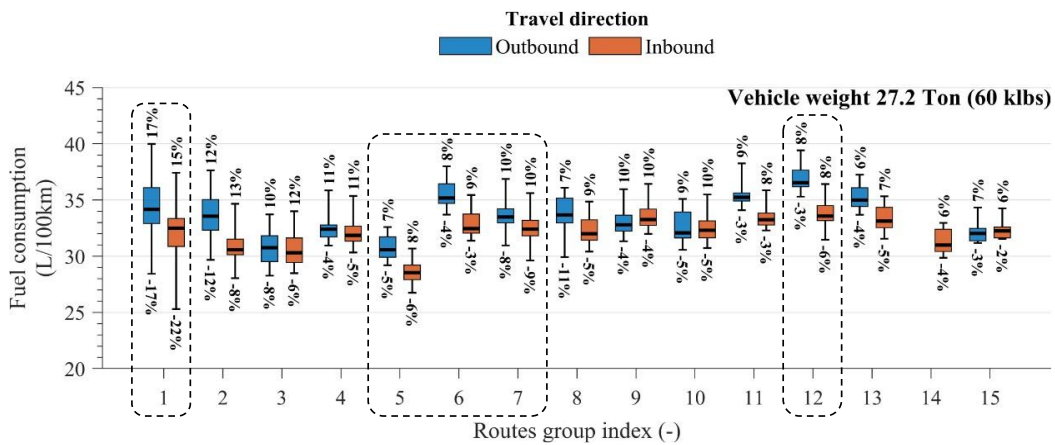

Figure 18: Fuel consumption variation of a conventional truck due to geographical and seasonal variations on different inbound and outbound route groups for a 27.2 Ton (60,000 lbs.) vehicle at Port of Savannah, GA.

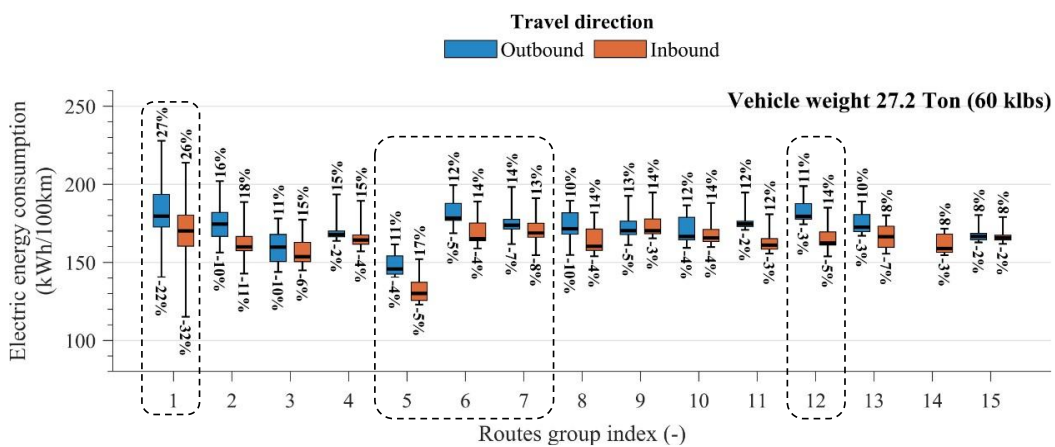

Figure 19: Electric energy consumption variation of an electric truck due to geographical and seasonal variations on different inbound and outbound route groups for a 27.2 Ton (60,000 lbs.) vehicle at Port of Savannah, GA.

Seasonal variations in weather conditions lead to fluctuations in energy consumption throughout the year, as previously discussed. To provide a detailed view of how weather differences affect travelling on the same route in various months, the monthly energy consumption (fuel and electricity) for conventional and electric trucks on route groups 5, 6, and 7 are presented in Figures 20 and 21, respectively. These figures show a clear seasonal trend in energy consumption, with vehicles requiring more energy to cover the same routes during the colder winter months compared to the warmer summer months. Additionally, the winter months exhibit higher variations in fuel and electricity consumption than the summer months. Take the case of route group 6 for instance, in December, the fuel and electricity consumption variations range from -14% to 16% and -24% to 26% of the month median consumption values, respectively. In contrast, July shows much lower variations, with fuel consumption ranging from -3% to 1% and electricity consumption from -2% to 1% of the month median consumption values. This difference can be attributed to the significant variation in average route temperatures during the winter months for cold, nominal, and hot days, while the average temperatures during the summer months remain relatively stable for cold, nominal, and hot days as already seen in Figure 14.

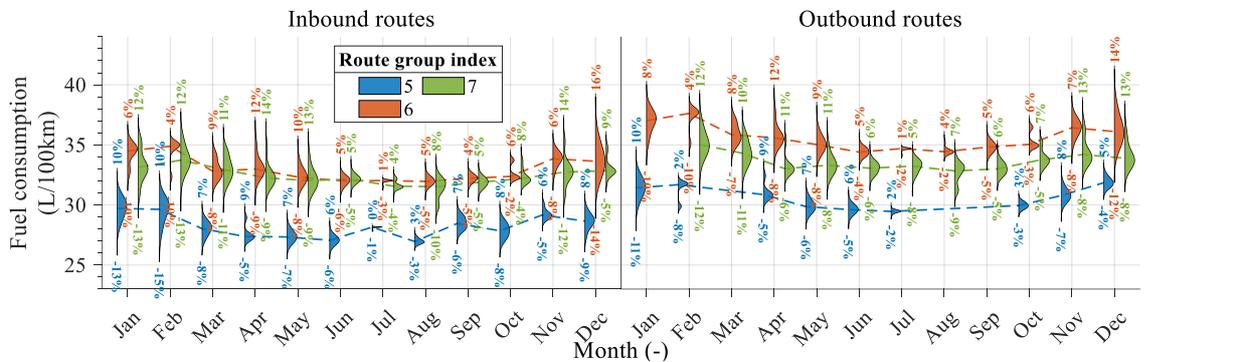

Figure 20: Month-wise fuel consumption variation of the conventional trucks on route groups 5, 6 and 7 due to differences in weather conditions in each month at Port of Savannah, GA.

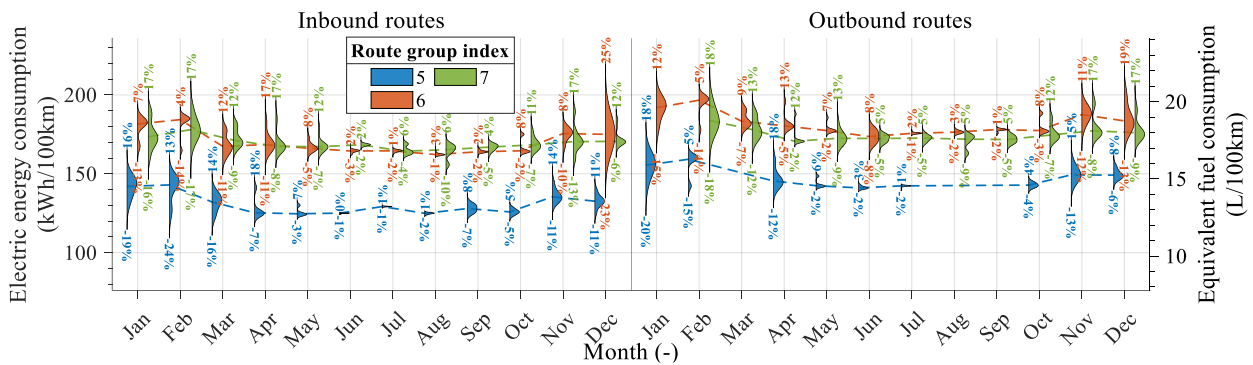

Figure 21: Month-wise electric energy consumption variation of electric trucks on route groups 5, 6 and 7 due to differences in weather conditions in each month at Port of Savannah, GA.

Figure 20 and 21 can be further utilized to evaluate the impact of geographical factors such as speed limit and elevation. As shown in Figure 17, route group 5 has the lowest average speed limit among route groups 5, 6, and 7. Consequently, route group 5 exhibits the lowest fuel and electricity consumption throughout the year. With the truck speed constrained by the road speed limit, vehicles travel the slowest on route group 5, reducing aerodynamic drag and rolling resistance, which are highly sensitive to speed. When comparing route groups 6 and 7, which are similar in all aspects except elevation change, it is observed that fuel consumption is lower on route group 7 than on route group 6 for both inbound and outbound trips; however, when analyzing the electric energy consumption of the electric truck on the same route groups, the negative elevation changes on inbound trips for route group 6 provide significant regeneration opportunities. This reduces the difference in electricity consumption between route groups 6 and 7. In contrast, this regenerative benefit is not available when traveling on the outbound trips on these route groups.

## *4.2 Variations in Auxiliary Power Consumption due to Seasonal Variability at Port of Savannah, GA*

The impact of seasonal variations on auxiliary power consumption for both conventional and electric trucks can be analyzed in detail. Figure 22 depicts the distribution of average auxiliary consumption in different months of the year, categorized by day type. For conventional trucks, auxiliary power consumption roughly follows an inverted U-shape, with the highest consumption observed during the hottest days of the summer months and the lowest during the coldest days of the winter months in the Port of Savannah region. Generally, auxiliary power consumption is lowest on days when the average route ambient temperature is around 20°C, as HVAC power consumption is minimal at this temperature.

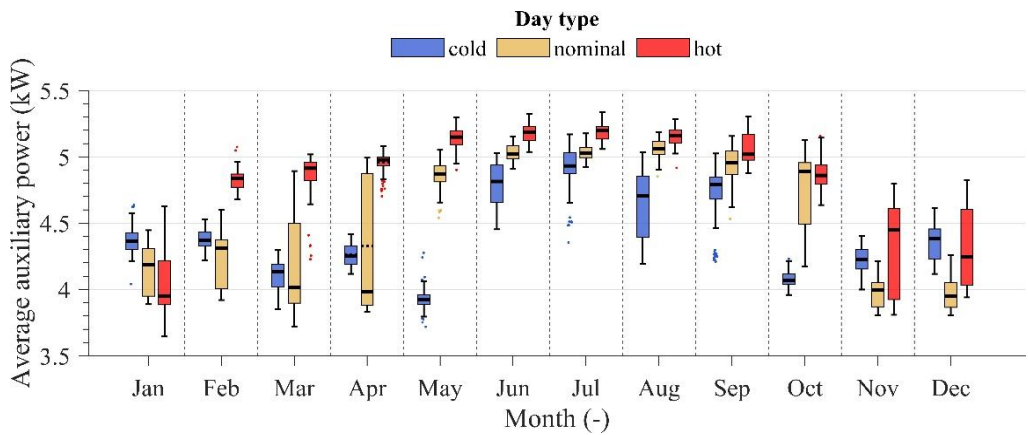

Figure 22: Conventional (diesel) truck average total auxiliary power consumption for different months and temperature day type at Port of Savannah, GA.

Thus, the overall trend in auxiliary power consumption for conventional trucks can be attributed to the HVAC system, which is significantly impacted by ambient temperature and is the major source of auxiliary power consumption, second only to the lubricant oil pump.

Similar observations are made for the electric truck auxiliary power consumption as illustrated in Figure 23. The major auxiliaries in the electric truck are the BTMS and HVAC, both of which are influenced by the ambient temperature. The auxiliary power consumption is lowest during the time of spring (March-May) and fall (September-October) where the temperatures are moderate such that both HVAC and BTMS systems are not required for maintaining the desired cabin and battery temperature during operation. While the higher auxiliary power consumption is observed during the coldest days of winter months as well as during the hottest days of summer months, although the maximum is for winter months.

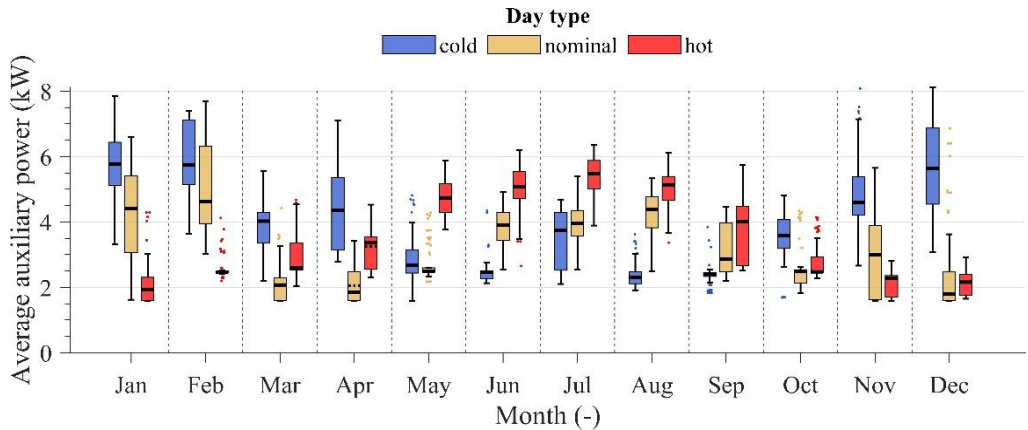

Figure 23: Electric truck average total auxiliary power consumption for different months and temperature day type at Port of Savannah, GA.

For the cold days in the month of January, the BTMS heater is used to warm up the battery and the HVAC system is used to heat up the cabin; however, the BTMS heater forms the major fraction (60-80%) of the total auxiliary power consumption during the coldest days of the year. Similarly, the BTMS chiller is required during the hot days of July to cool down the battery and has the biggest contribution to total auxiliary consumption. Again, HVAC contribution comes in second after BTMS chiller during the hottest time of the year.

## *4.3 Drayage Truck Energy Consumption Variability Across Distinct Ports*

A similar analysis was conducted for two additional ports with significantly different weather and geographical conditions: the Port of Houston, TX, and the Port of Seattle/Tacoma, WA. These ports were selected to capture the impact of contrasting climates−humid subtropical in Houston and cold maritime in Seattle/Tacoma, WA− on vehicle energy consumption. Appendix B provides an overview of the simulation results for these two ports, including the routes information, the variation in energy consumption for both conventional and electric trucks, as well as the seasonal variation in auxiliary power consumption across different months. Comparing the energy consumption results of a 27.2 Ton (60,000 lbs.) electric truck from all three ports offers valuable insights into the variations in energy consumption that electric drayage trucks are likely to encounter under different environmental and operational conditions and are tabulated in Table 5 and 6.

Table 5: Mean plus/minus one standard deviation of electric energy consumption (kWh/100km) of 27.2 Ton (60,000 lbs) electric drayage trucks on inbound trips at three distinct ports. The energy consumption values are grouped by route distance and seasons.

| | | Season | | | | Port |
|---|---|---|---|---|---|---|
| | | Spring | Summer | Autumn | Winter | |
| Route distance | Less than 80 km | 165.5 ± 16 | 160.6 ± 16 | 156.1 ± 20 | 170.7 ± 21 | Savannah, GA |
| | | 169.4 ± 12 | 166.1 ± 9 | 167.3 ± 11 | 180.8 ± 21 | Houston, TX |
| | | 197.1 ± 20 | 191.9 ± 20 | 198.5 ± 17 | 205.3 ± 20 | Seattle/Tacoma, WA |
| | Between 80 and 240 km | 163.7 ± 17 | 160.6 ± 13 | 163.7 ± 14 | 170.2 ± 18 | Savannah, GA |
| | | 158.4 ± 10 | 155.5 ± 8 | 157.4 ± 10 | 169.4 ± 18 | Houston, TX |
| | | 196.3 ± 15 | 184.8 ± 19 | 191.6 ± 16 | 203.1 ± 18 | Seattle/Tacoma, WA |
| | Between 240 and 400 km | 164.3 ± 7 | 159.1 ± 4 | 160.8 ± 11 | 170.9 ± 10 | Savannah, GA |
| | | 161.9 ± 12 | 157.2 ± 10 | 159.9 ± 13 | 167.7 ± 16 | Houston, TX |
| | | 194.4 ± 11 | 188 ± 11 | 195.3 ± 10 | 202.3 ± 12 | Seattle/Tacoma, WA |

Table 6: Mean plus/minus one standard deviation of electric energy consumption (kWh/100km) of 27.2 Ton (60,000 lbs) electric drayage trucks on outbound trips at three distinct ports. The energy consumption values are grouped by route distance and seasons.

| | | Season | | | | Port |
|---|---|---|---|---|---|---|
| | | Spring | Summer | Autumn | Winter | |
| Route distance | Less than 80 km | 185.3 ± 15 | 169.4 ± 16 | 185.1 ± 13 | 191.7 ± 25 | Savannah, GA |
| | | 177.4 ± 14 | 170.7 ± 9 | 173.7 ± 11 | 187.4 ± 20 | Houston, TX |
| | | 211.3 ± 25 | 196.6 ± 23 | 202.6 ± 25 | 214.4 ± 25 | Seattle/Tacoma, WA |
| | Between 80 and 240 km | 171.6 ± 15 | 170.8 ± 7 | 172.0 ± 12 | 177.0 ± 16 | Savannah, GA |
| | | 167.1 ± 13 | 161.3 ± 8 | 166.6 ± 11 | 175.8 ± 19 | Houston, TX |
| | | 212.6 ± 26 | 204.3 ± 25 | 211.3 ± 25 | 216.0 ± 23 | Seattle/Tacoma, WA |
| | Between 240 and 400 km | 173.4 ± 12 | 170.3 ± 8 | 174 ± 9 | 182.6 ± 10 | Savannah, GA |
| | | 172.3 ± 14 | 170.2 ± 11 | 172.2 ± 11 | 179.9 ± 16 | Houston, TX |
| | | 199.8 ± 11 | 193.3 ± 9 | 197.0 ± 9 | 208.4 ± 13 | Seattle/Tacoma, WA |

The results from the three ports indicate that the overall average energy consumption is highest across the board at the Port of Seattle/Tacoma, WA, primarily due to the region's relatively colder climate which leads to higher road load (as a consequence of low temperatures and high air density) along with higher auxiliary power consumption, primarily from the BTMS and HVAC for heating. In contrast, the other two ports exhibit comparable energy consumption levels throughout the year, which can be attributed to their similar humid subtropical climates. The winter season presents the greatest challenge across all three regions, with the highest average energy consumption and the largest variability, as evidenced by high standard deviation values. Conversely, the summer season has the lowest energy consumption across the three ports, while spring and autumn show similar energy consumption patterns. The previously observed trend of inbound routes requiring less energy than outbound routes remain consistent across the other two ports as well. This comparative analysis highlights the limitations of using a simplified, fixed energy consumption rate for vehicles, as employed in some prior studies, which can result in significantly inaccurate conclusions if the energy consumption rates does not consider the dependence on geography and climate of the region.

Further inferences on driving range and payload capability of electric drayage trucks at these three ports can be made from these results. In general, for the same battery pack capacity, the trucks in the Port of

Seattle/Tacoma, WA will have the lowest range out of all three ports and may not be able to complete longer trips on a single charge, as the vehicle energy consumption is close to 250 kWh/100km on most of the routes during harsh winter months for a 27.2 Ton (60,000 lbs.) truck. Similarly, in the Port of Houston, TX, where on few occasions during the months of January and February the temperatures go below freezing, some trips may not be completed on a single charge. However, during mild months the average energy consumption is in the range of 160-200 kWh/100km across the board for a 27.2 Ton truck. Considering these energy consumption values, an 800 kWh battery pack can provide enough energy to complete most of the routes inside a 400 km (~250 mile) distance from the port in these regions. Conversely, if the driving range is to be met in the harsh weather conditions or a longer route needs to be travelled, then the payload needs to be reduced, or the battery pack size must be increased which will nevertheless result in reduction of payload capacity and increase in vehicle cost.

## *4.4 Implication of Seasonal Variations on Vehicle Monthly Energy Consumption at Port of Savannah, GA*

As shown earlier, seasonal weather variations significantly influence vehicle energy consumption across different routes. To further understand how these weather variations impact the monthly energy consumptions of drayage trucks, the operation at Port of Savannah, GA is considered again. The case of drayage trucks with 60,000 miles (96,560 km) of annual vehicle miles travelled (VMT) is analyzed. In this scenario, trucks are expected to drive 5,000 miles (8,046 km) each month around the port. Month wise daily vehicle trip itineraries are constructed by randomly sampling the routes, weather conditions and cargo weights. In addition, due to the differences in curb weight of the conventional and electric truck, as well as the difference in the curb weight of the different battery sized electric truck, not all trips and vehicle weight are possible for both the vehicle types.

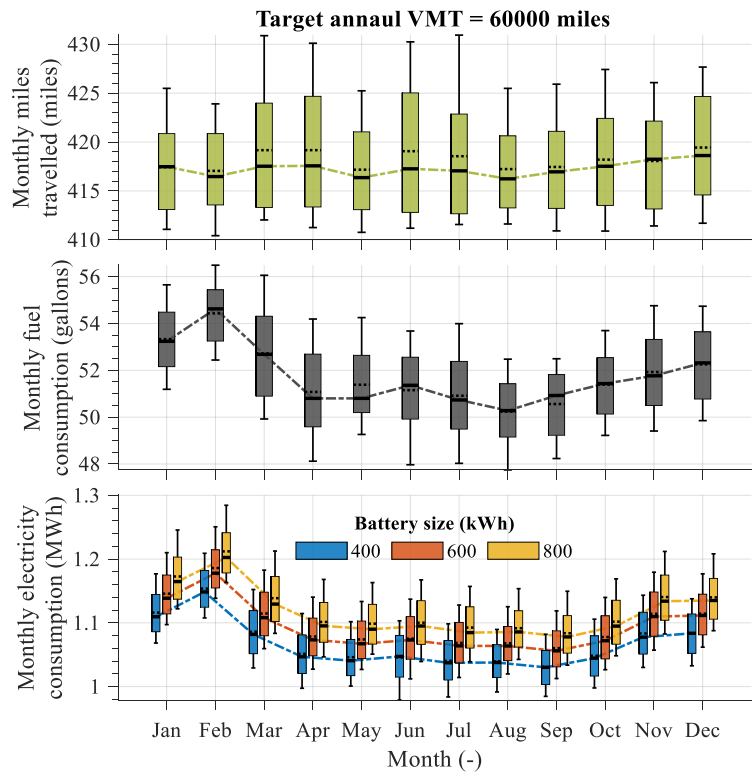

Figure 24: Distribution of monthly miles traveled by trucks for the case of 60,000 annual VMT target (top); Monthly fuel consumption distribution of diesel trucks (middle), and Monthly electricity consumption distribution of electric trucks considering three battery sizes (bottom).

The top part of Figure 24 shows the distribution of simulated monthly VMT for 40 trucks in this scenario, indicating that the simulated trucks indeed travel an average of 5,000 miles (8046 km) each month. The middle and bottom parts of the figure provide the monthly fuel and electricity consumption for diesel and electric trucks, respectively. Energy consumption, for both fuel and electricity, is higher during the colder winter months, peaking in February. However, the high temperatures during the summer month of July do not have a similar effect on energy consumption, with monthly energy consumption remaining relatively consistent from April to October. Specifically for electric trucks, it is noted that trucks with larger battery sizes result in higher monthly electricity consumption owing to the extra weight the battery pack adds to the total vehicle weight.

These results indicate that seasonal weather variations can lead to significant differences in monthly energy consumption for both conventional and electric trucks. Consequently, the energy requirements for refueling conventional trucks and recharging electric trucks are also affected by these seasonal changes. This difference in monthly fuel and electricity consumption will be particularly pronounced in regions that experience substantial weather fluctuations throughout the year. As a result, these seasonal weather variations are likely to impact vehicle operational costs. Subsequently, the total cost of ownership (TCO) of the truck considering the variation in operational costs due to regional geographical and seasonal factors can be determined using a TCO model developed in Sun et al. (2024).

## 5. Conclusions

The adoption of battery electric trucks in the drayage sector is a crucial step toward the electrification of port operations in the United States; however, the varied geographical characteristics and seasonal weather conditions of different ports necessitates that each port/region be analyzed individually. This requires a tailored techno-economic assessment of the benefits and challenges in electrifying the current conventional diesel-powered drayage fleet at each port. To make a reliable and accurate techno-economic assessment, a fundamental step is the development of a detailed bottom-up simulation-based framework for the energy consumption estimation of heavy-duty vehicles. The models developed in this study are proposed to replace the more commonly used lumped energy consumption values, e.g., kWh/km or l/100 km, that are currently used in most techno-economic analyses, with the goal of providing more accurate and region-specific energy consumption estimations for drayage fleets operating in different climates. In this work, vehicle simulators for both conventional and electric trucks, which account for geographical and weather characteristics impacting the road load as well as the auxiliaries, are developed to determine vehicle energy consumption. The variation in energy consumption obtained from the simulations is validated with the limited available real-world experimental data. Due to the limited availability of experimentally collected operational data demonstrating seasonal variations in energy consumption for electric trucks, a bottom-up approach to vehicle energy consumption modeling has been employed. This limited availability of data also hinders from performing a rigorous model validation for the vehicle energy consumption model of the electric truck and is currently a limitation of this work. Nonetheless, efforts are also being made to collaborate with industry partners to acquire top-down data from in-operation electric trucks to corroborate the findings presented in this study.

The results reveal that seasonal and geographical variations significantly impact energy consumption for both powertrain types. Ambient temperature notably affects aerodynamic drag and rolling resistance, as it directly influences air density and tire temperature. Seasonal variations in ambient temperature also profoundly affect auxiliary power consumption, particularly for HVAC systems in both truck types and the BTMS in electric trucks. During harsh cold weather conditions, the BTMS can account for 60-80% of the overall auxiliary power consumption in an electric truck. Geographical characteristics, such as elevation changes and average route speed limits, also significantly influence energy consumption. Routes with negative elevation changes favor electric trucks, as they can harness energy through regenerative braking, reducing power consumption drastically.

Although the simulation of a single route of roughly 120 miles (~193 km) for a single vehicle takes less than 45 seconds to run on an average CPU (e.g., AMD Ryzen 9 5900HX), running millions of such simulation can quickly become a bottleneck to perform the kind of analysis presented in this work and future techno-economic analyses. Utilizing high-performance computational resource, such as those available at the Ohio

Supercomputer Center, drastically reduces the total time required for carrying out the analysis presented in this paper, by parallelizing the simulations on multiple CPU cores. However, it still takes about half a day to complete all the simulations for a port using 200 CPU cores on the Supercomputer. To eliminate this dependence on the use of such high-performance computing facilities and make the simulator and models easily accessible to all, there are efforts currently underway to develop a reduced-complexity model that leverages regression-based machine learning approaches to estimate vehicle energy consumption based on limited route and vehicle information. Using this reduced-complexity model will be significantly faster than running all the simulations and facilitate an easy integration of the accurate energy consumption model into techno-economic tools developed by various researchers.

Future work will evaluate a broader range of critical U.S. ports to provide valuable insights for decision-makers regarding the region-specific adoption of electric trucks in drayage fleets, ensuring that current port operations and logistics remain unaffected. Additionally, the forthcoming study will investigate various battery chemistries to gain a comprehensive understanding of which lithium-ion batteries are most suitable for heavy-duty applications, with a particular emphasis on drayage operations. This research will incorporate detailed semi-empirical battery aging models, accounting for both calendar and cyclic aging, to assess the necessity of battery replacements in electric drayage trucks. Furthermore, by utilizing the detailed electric truck energy consumption model, along with battery degradation models and drayage truck activity modeling, the total cost of ownership will be analyzed.

## Author contributions

**Ankur Shiledar:** Conceptualization, Data curation, Formal analysis, Investigation, Methodology, Software, Validation, Visualization, Writing – original draft. **Manfredi Villani:** Formal analysis, Investigation, Project administration, Writing – original draft, Writing - Reviewing and Editing. **Joseph N.E. Lucero:** Data curation, Formal analysis, Investigation, Software, Validation, Writing – original draft. **Ruixiao Sun:** Writing - Reviewing and Editing. **Vivek A. Sujan:** Conceptualization, Funding acquisition, Project administration, Supervision. **Simona Onori:** Project administration, Supervision, Writing - Reviewing and Editing. **Giorgio Rizzoni:** Project administration, Supervision, Writing - Reviewing and Editing.

## Acknowledgments

This study received funding and support from the Department of Energy Vehicle Technologies Office through the award WBS 7.2.0.502 / FWP CEVT442, administered by the Oak Ridge National Laboratory's National Transportation Research Center.

Giuliano, Genevieve, Maged Dessouky, Sue Dexter, Jiawen Fang, Shichun Hu, and Marshall Miller. 2021. "Heavy-Duty Trucks: The Challenge of Getting to Zero." *Transportation Research Part D: Transport and Environment* 93 (April). https://doi.org/10.1016/j.trd.2021.102742.

Giuliano, Genevieve, and Thomas O'Brien. 2007. "Reducing Port-Related Truck Emissions: The Terminal Gate Appointment System at the Ports of Los Angeles and Long Beach." *Transportation Research Part D: Transport and Environment* 12 (7): 460–73. https://doi.org/10.1016/j.trd.2007.06.004.

"Global Commercial Drive To Zero Program — Global Memorandum of Understanding on Zero-Emission Medium- and Heavy-Duty Vehicles." n.d. Accessed April 11, 2024. https://globaldrivetozero.org/mou-nations/#.

Golbasi, Onur, and Elif Kina. 2022. "Haul Truck Fuel Consumption Modeling under Random Operating Conditions: A Case Study." *Transportation Research Part D: Transport and Environment* 102 (January). https://doi.org/10.1016/j.trd.2021.103135.

Hall, Dale, and Nic Lutsey. 2019. "Estimating the Infrastructure Needs and Costs for the Launch of Zero-Emission Trucks." *ICCT The International Council on Clean Transportation*, August. https://theicct.org/wp-content/uploads/2021/06/ICCT_EV_HDVs_Infrastructure_20190809.pdf.

Howell, J P. 2014. "Aerodynamic Drag in a Windy Environment." In *The International Vehicle Aerodynamics Conference*, edited by Holywell Park, 19–30. Woodhead Publishing. https://doi.org/10.1533/9780081002452.1.19.

Hunter, Chad, Michael Penev, Evan Reznicek, Jason Lustbader, Alicia Birky, and Chen Zhang. 2021. "Spatial and Temporal Analysis of the Total Cost of Ownership for Class 8 Tractors and Class 4 Parcel Delivery Trucks." www.nrel.gov/publications.

Hyttinen, Jukka, Matthias Ussner, Rickard Österlöf, Jenny Jerrelind, and Lars Drugge. 2022. "Effect of Ambient and Tyre Temperature on Truck Tyre Rolling Resistance." *International Journal of Automotive Technology* 23 (6): 1651–61. https://doi.org/10.1007/s12239-022-0143-6.

Hyttinen, Jukka, Matthias Ussner, Rickard Österlöf, Jenny Jerrelind and Lars Drugge . 2023. "Truck Tyre Transient Rolling Resistance and Temperature at Varying Vehicle Velocities - Measurements and Simulations." *Polymer Testing* 122 (May). https://doi.org/10.1016/j.polymertesting.2023.108004.

Khuntia, Satvik, Athar Hanif, Somendra Pratap Singh, and Qadeer Ahmed. 2022. "Control Oriented Model of Cabin-HVAC System in a Long-Haul Trucks for Energy Management Applications." In *SAE Technical Papers*. SAE International. https://doi.org/10.4271/2022-01-0179.

## Definitions/Abbreviations

| | |
|---|---|
| BETs | battery electric trucks |
| BTMS | battery thermal management system |
| ECM | equivalent circuit model |
| GHG | greenhouse gases |
| HD | heavy duty |
| HDV | heavy duty vehicle |
| HVAC | heating, ventilation, and air conditioning |
| LDV | light duty vehicle |
| mEDM | modified enhanced driver model |
| OD | origin-destination |
| RR | rolling resistance |
| SOC | state of charge |
| TCO | total cost of ownership |
| VMT | vehicle miles travelled |
| ZEV | zero emissions vehicle |

# Appendix A

A separate dataset containing the daily average temperature of all the major cities along all the identified routes is created using WUnderground ("Local Weather Forecast, News and Conditions | Weather Underground," n.d.) temperature data. Next, for each route, for the corresponding month it belongs to, a route temperature vector is constructed for all the days of the month using equation A1. The elements of this route temperature vector are the average temperature (in units of K) of all $N$ major cities along the route on that day. For each route, $D$ count of such route temperature vectors are created, where $D$ equals the total days for the corresponding month.

$$T_{route,i} = [T_{1,i} \quad \cdots \quad T_{N,i}]^T, \qquad i = 1, \ldots, D \tag{A1}$$

This is followed by calculating the representative route temperature for the day as follows:

$$\overline{T_{route,\iota}} = \sqrt{\frac{1}{N}\sum_{k=1}^{N} T_{k,i}^2} \tag{A2}$$

Using this representative route temperature, the hottest, the coldest and nominal (median) temperature dates of the months for the concerned route are determined as follows:

$$i_{hot} = \underset{i}{\mathrm{argmax}}\, \overline{T_{route,\iota}} \tag{A3a}$$

$$i_{cold} = \underset{i}{\mathrm{argmin}}\, \overline{T_{route,\iota}} \tag{A3b}$$

$$i_{nominal} = \underset{i}{\mathrm{argmin}}\, \left(median\bigl(\overline{T_{route,1:D}}\bigr) - \overline{T_{route,\iota}}\right)^2 \tag{A3c}$$

Here, $median\bigl(\overline{T_{route,1:D}}\bigr)$ is the median representative temperature of the route for the given month. The route and weather data for an example inbound and outbound route are shown in Figure A1. The dates identified for cold, nominal, and hot conditions for the chosen inbound route in May of 2021 are, 13, 19, and 27, respectively. Similarly, for the selected outbound route in the month of January 2021, the identified cold, nominal and hot conditions dates are 29, 15, and 26, respectively.

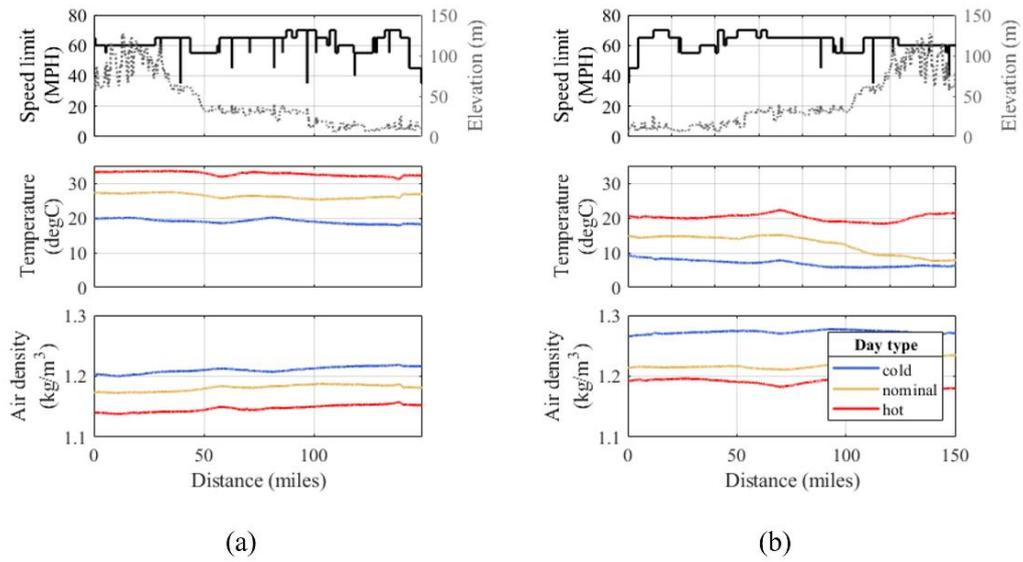

Figure A1: Route and weather data for (a) a particular inbound route in the month of May and (b) a particular outbound route in the month of January.

# Appendix B

## *Port of Houston, TX*

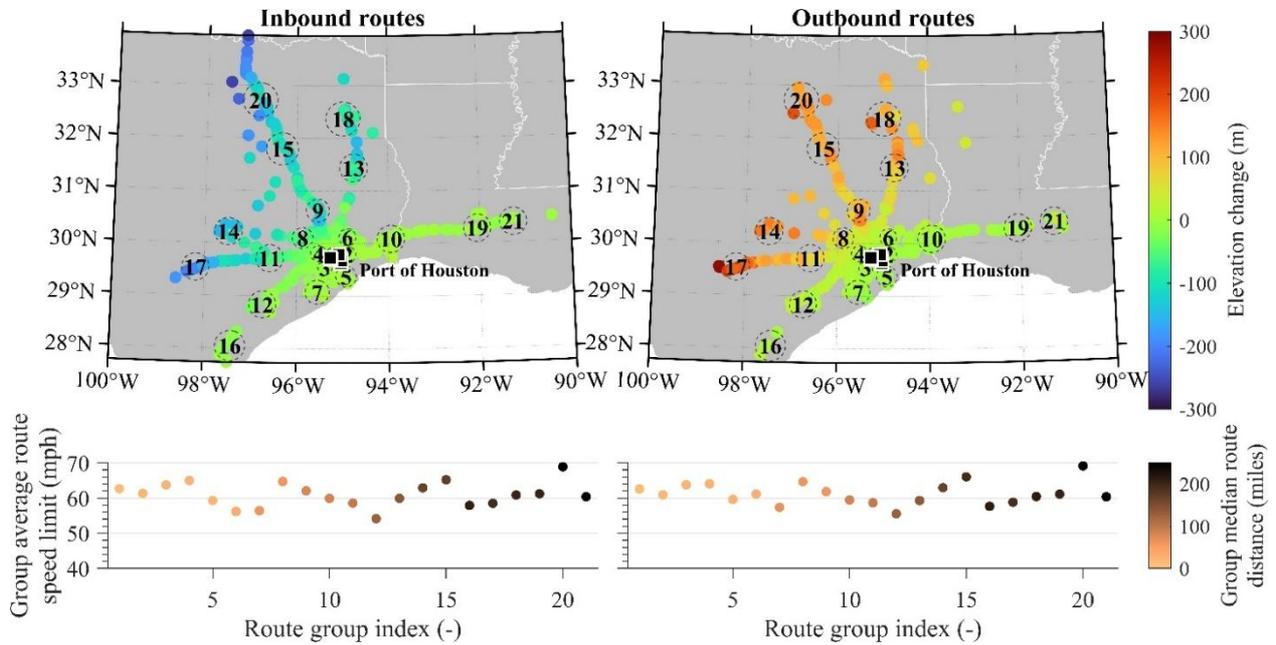

Figure B1: Inbound (left) and outbound (route) route grouping for analyzing the impact of geographical and weather variation on energy consumption in HD trucks at Port of Houston, TX. Figure formatting same as Figure 17.

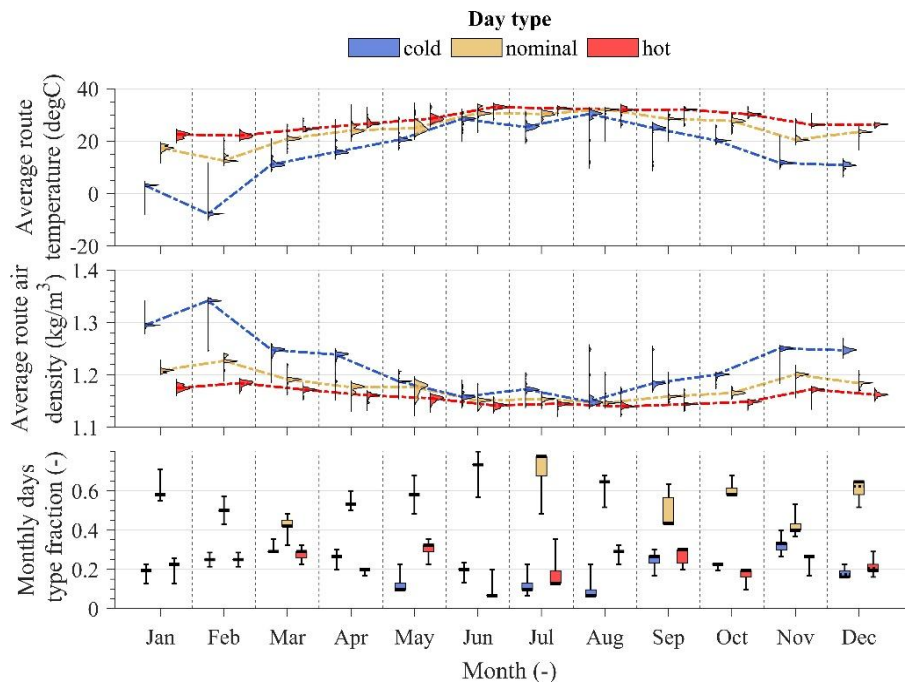

Figure B2: Monthly weather condition variations in the region around the Port of Houston, TX.

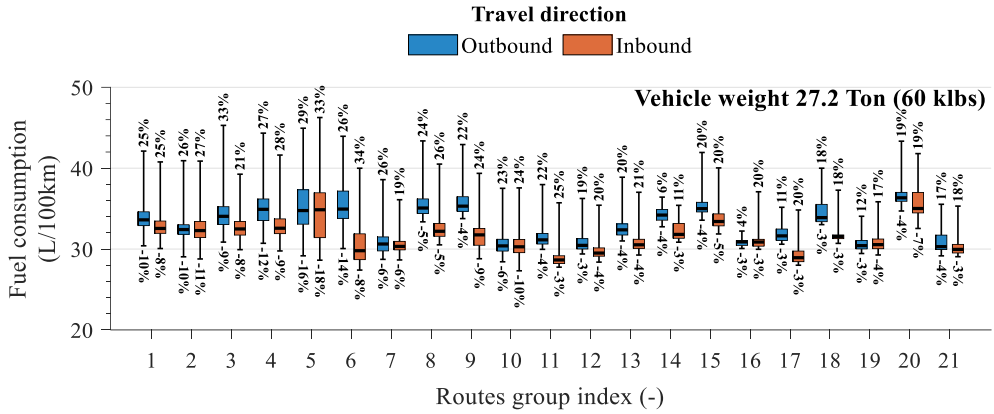

Figure B3: Fuel consumption variation of a conventional truck due to geographical and seasonal variations on different inbound and outbound route groups for a 60,000 lbs. vehicle at Port of Houston, TX.

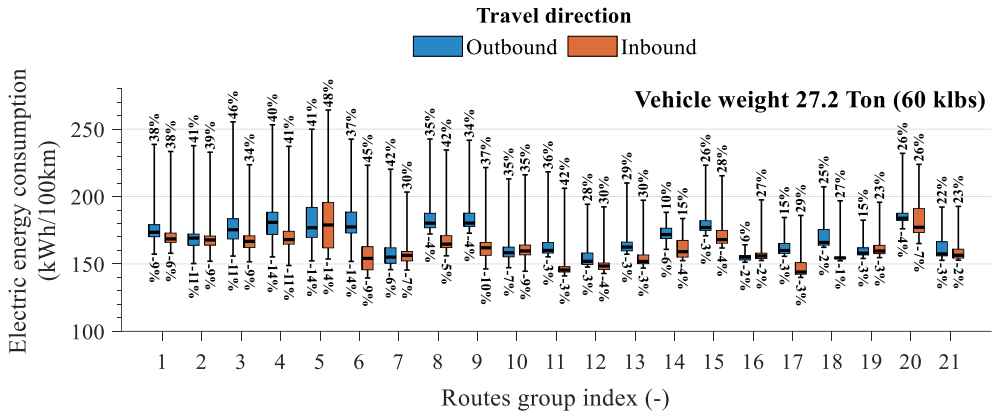

Figure B4: Electric energy consumption variation of an electric truck due to geographical and seasonal variations on different inbound and outbound route groups for a 60,000 lbs. vehicle at Port of Houston, TX.

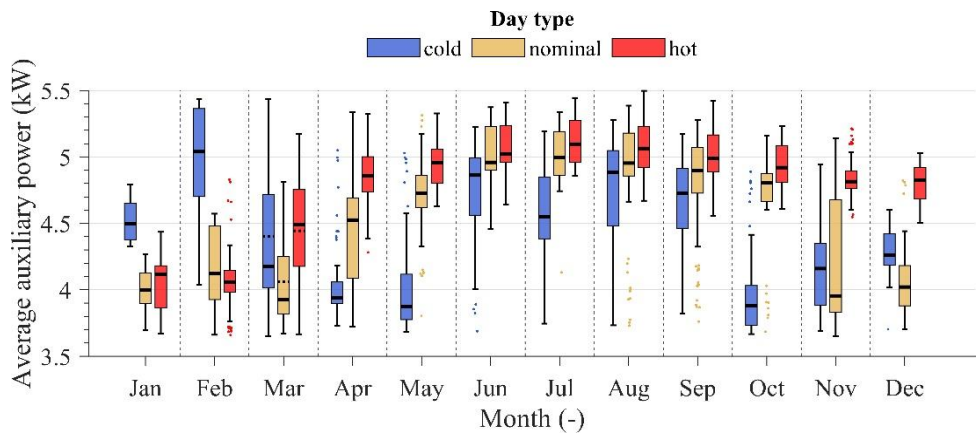

Figure B5: Conventional (diesel) truck average total auxiliary power consumption for different months and temperature day type at Port of Houston, TX.

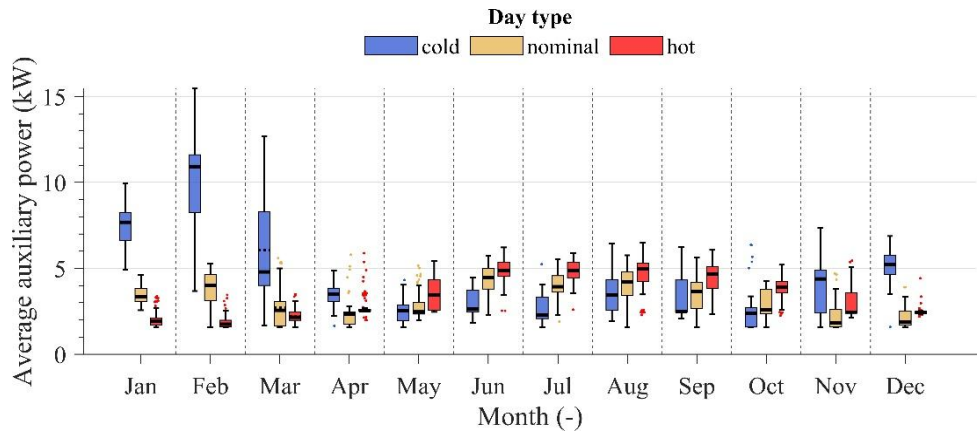

Figure B6: Electric truck average total auxiliary power consumption for different months and temperature day type at Port of Houston, TX.

## Port of Seattle/Tacoma, WA

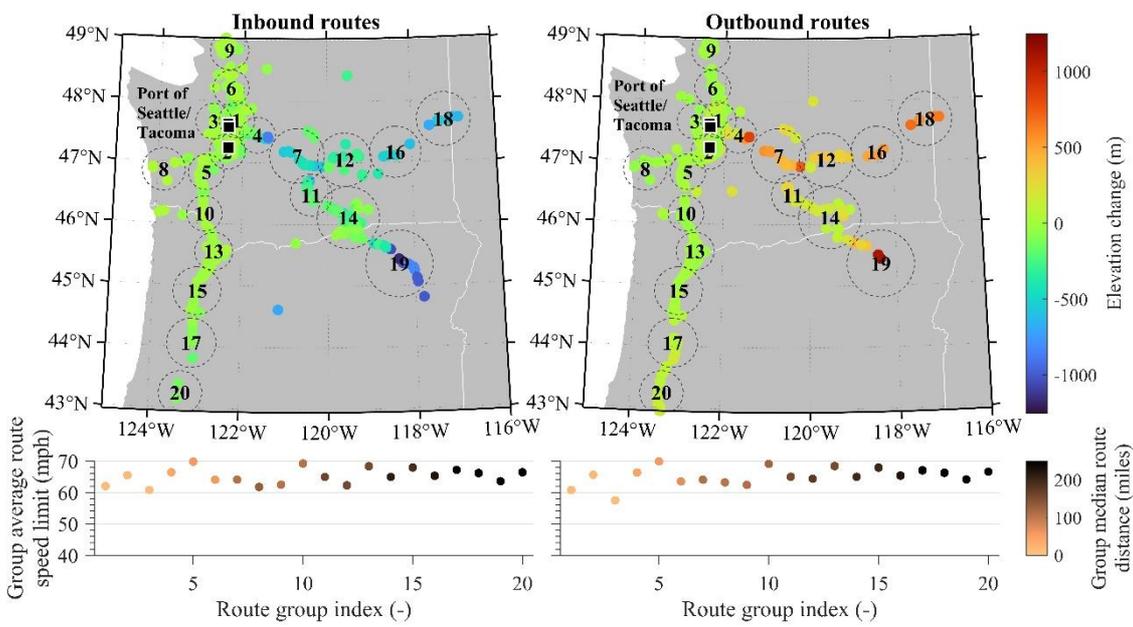

Figure B7: Inbound (left) and outbound (route) route grouping for analyzing the impact of geographical and weather variation on energy consumption in HD trucks at Port of Seattle/Tacoma, WA. Figure formatting same as Figure 17.

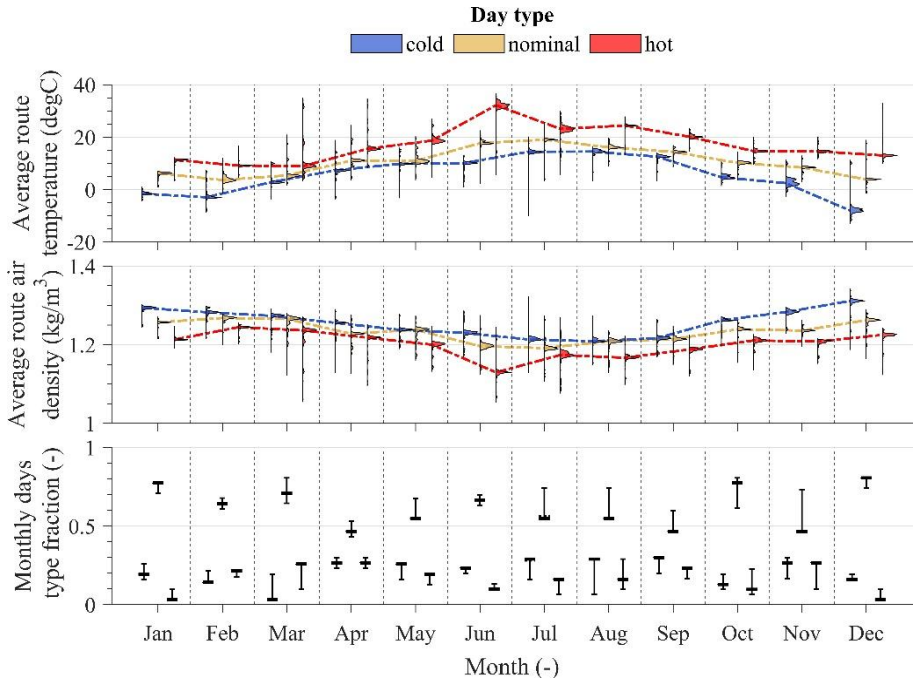

Figure B8: Monthly weather condition variations in the region around the Port of Seattle/Tacoma, WA.

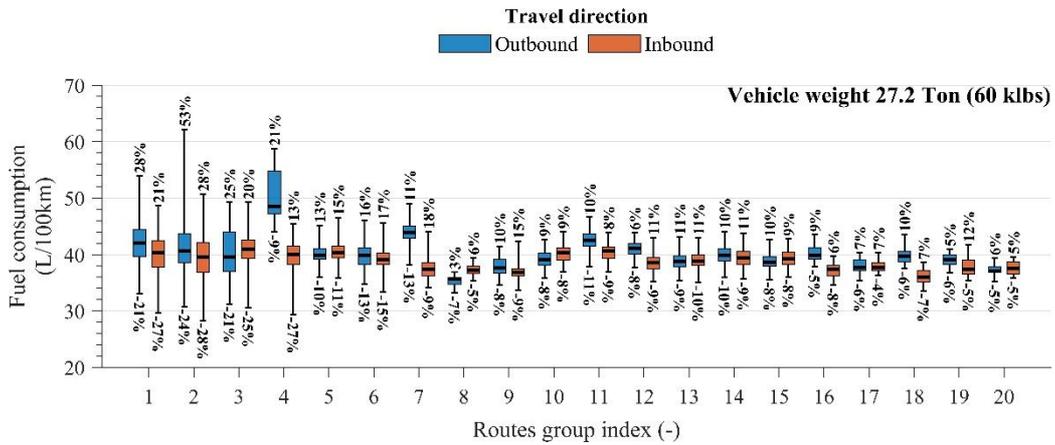

Figure B9: Fuel consumption variation of a conventional truck due to geographical and seasonal variations on different inbound and outbound route groups for a 60,000 lbs. vehicle at Port of Seattle/Tacoma, WA.

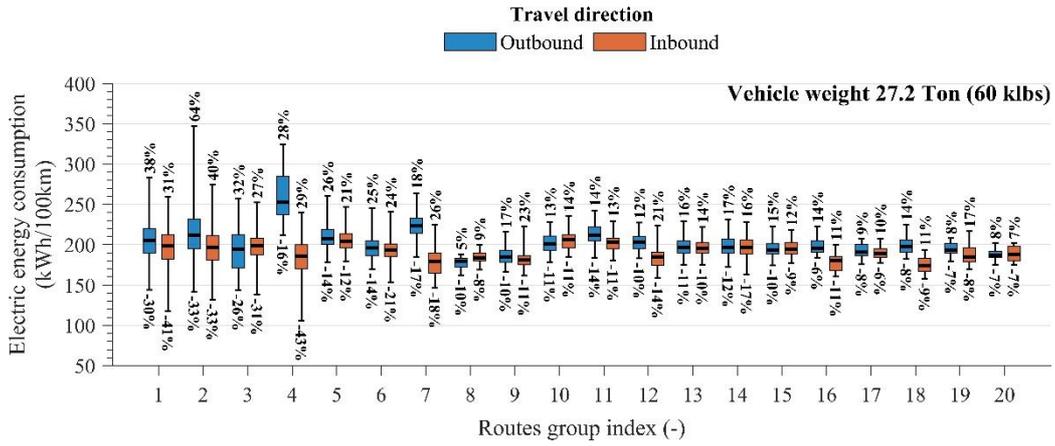

Figure B10: Electric energy consumption variation of an electric truck due to geographical and seasonal variations on different inbound and outbound route groups for a 60,000 lbs. vehicle at Port of Seattle/Tacoma, WA.

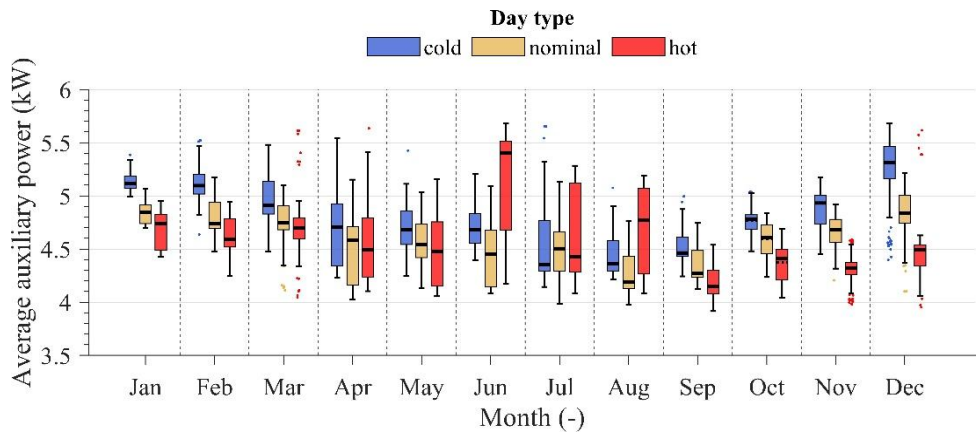

Figure B11: Conventional (diesel) truck average total auxiliary power consumption for different months and temperature day type at Port of Seattle/Tacoma, WA.

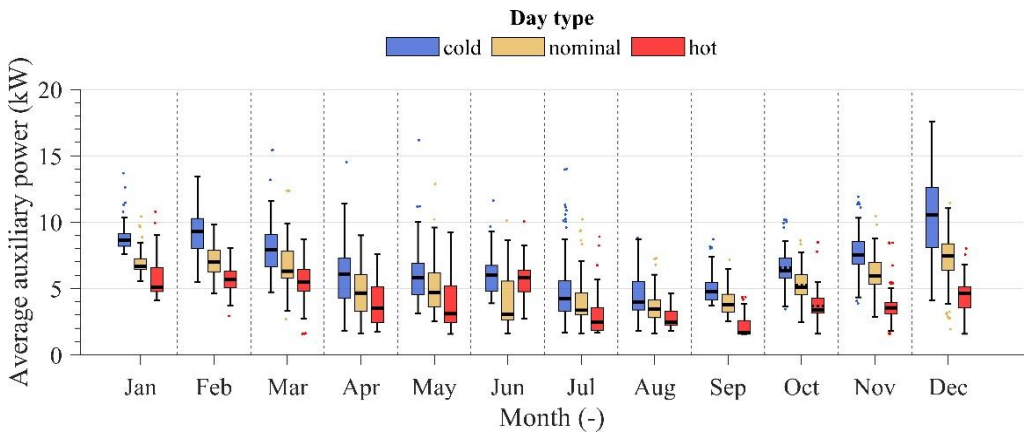

Figure B12: Electric truck average total auxiliary power consumption for different months and temperature day type at Port of Seattle/Tacoma, WA.